\documentclass[11pt, a4paper]{article}

\usepackage[margin=1in]{geometry}
\usepackage{setspace}
\usepackage[
backend=biber,
style=alphabetic,
citestyle=alphabetic,
maxalphanames=5,
maxcitenames=99,
mincitenames=98,
maxbibnames=99,
giveninits=true,
]{biblatex}

\usepackage[T1]{fontenc}
\usepackage{lmodern}
\usepackage{amsmath,amssymb}
\usepackage{amsthm}
\usepackage{thm-restate}
\usepackage{enumitem}
\usepackage[dvipsnames]{xcolor}
\usepackage{xspace,mathtools}
\usepackage{hyperref}
\usepackage[capitalise,nameinlink,noabbrev]{cleveref}

\theoremstyle{plain}
\newtheorem{theorem}{Theorem}[section]
\newtheorem{lemma}[theorem]{Lemma}
\theoremstyle{definition}
\newtheorem{definition}[theorem]{Definition}

\crefname{ineq}{inequality}{inequalities}
\Crefname{ineq}{Inequality}{Inequalities}
\crefname{cons}{constraint}{constraints}
\Crefname{cons}{Constraint}{Constraints}

\newenvironment{proofof}[1]{\begin{proof}[Proof of #1]}{\end{proof}}

\usepackage{tikz}
\usetikzlibrary{calc,fit,positioning,backgrounds,shapes.geometric,shapes.symbols,decorations.pathmorphing}
\usepackage{hyperref}
\hypersetup{colorlinks=true,linkcolor=blue!60!black,citecolor=blue!60!black,urlcolor=blue!60!black}

\colorlet{augcol}{red}
\definecolor{badcol}{RGB}{150,0,40}
\definecolor{optcol}{RGB}{0,140,60}
\definecolor{compcol}{RGB}{214,228,245}
\definecolor{spdcol}{RGB}{30,80,160}

\tikzset{
  act/.style={rectangle,fill=black,minimum size=5.4pt,inner sep=0pt},
  inact/.style={circle,draw=black,fill=white,line width=0.8pt,minimum size=6pt,inner sep=0pt},
  dot/.style={circle,fill=black!65,minimum size=4pt,inner sep=0pt},
  fo/.style={line width=1.3pt,black},
  family edge/.style={draw=red,thick},
  twin edge/.style={draw=red,thick,dashed},
  cross edge/.style={draw=red,thick,decorate,decoration={zigzag,segment length=4pt,amplitude=1.2pt}},
  inner edge/.style={draw=red,thick,densely dotted},
  aug/.style={family edge},
  twn/.style={twin edge},
  crs/.style={cross edge},
  inn/.style={inner edge},
  bad/.style={line width=1.3pt,badcol},
  opte/.style={line width=1.1pt,optcol,dash pattern=on 4pt off 2.5pt},
  anye/.style={line width=1.3pt,black!35},
  comp/.style={ellipse,fill=compcol,draw=compcol!65!black,line width=0.5pt,inner sep=0pt},
  spd/.style={font=\scriptsize,text=spdcol,inner sep=1pt},
  nl/.style={font=\small},
  lab/.style={font=\small},
  slab/.style={font=\footnotesize,text=black!70},
  elab/.style={font=\footnotesize,inner sep=1.5pt},
  pcap/.style={font=\footnotesize,text width=4.6cm,align=center,anchor=north},
  setb/.style={draw=black!65,thick},
  hl/.style={fill=orange!30},
  regn/.style={ellipse,draw=black!70,thick,fill=gray!12,inner sep=2.5pt,minimum width=1.0cm,minimum height=0.62cm,font=\small},
  regw/.style={ellipse,draw=black!70,thick,fill=white,inner sep=2.5pt,minimum width=1.0cm,minimum height=0.62cm,font=\small},
}

\newcommand{\OPT}{\mathrm{OPT}}

\newcommand{\ALG}{\mathrm{ALG}}

\newcommand{\twoecss}{\textrm{2ECSS}\xspace}
\newcommand{\kecss}{\mathrm{k}\textrm{ECSS}\xspace}

\newcommand{\bstar}{\beta^\star}

\newcommand{\Atw}{A^{\mathrm{twin}}}
\newcommand{\Amx}{A^{\mathrm{mixed}}}
\newcommand{\Ain}{A^{\mathrm{inner}}}
\newcommand{\Acr}{A^{\mathrm{cross}}}

\newcommand{\cT}{\mathcal{T}}
\newcommand{\cW}{\mathcal{W}}
\newcommand{\Wtw}{\cW^{\mathrm{twin}}}
\newcommand{\Wmx}{\cW^{\mathrm{mixed}}}
\newcommand{\region}{\mathrm{ex}}
\newcommand{\parent}{\mathrm{p}}
\newcommand{\children}{\mathrm{ch}}
\newcommand{\witness}{\mathrm{wit}}
\newcommand{\nactive}{q}

\newcommand{\C}{\mathcal{C}}

\newcommand{\rc}{{\bar c}}

\newcommand{\bone}[1]{\mathbf{1}[#1]}

\newcommand{\eps}{\varepsilon}
\newcommand{\sm}{\setminus}
\newcommand{\sse}{\subseteq}
\newcommand{\ssne}{\subsetneq}

\newcommand{\yvar}[1]{{y^{}_{#1}}}
\newcommand{\evar}[1]{{e^{}_{#1}}}
\newcommand{\fvar}[1]{{f^{}_{#1}}}

\newcommand{\APX}{\mathsf{APX}}
\newcommand{\NP}{\mathsf{NP}}
\newcommand{\Pclass}{\mathsf{P}}

\newcommand{\Rp}{\ensuremath{\mathbb R}_{\geq 0}}

\newcommand{\del}{\delta}

\newcommand{\FigureClusters}{%
\begin{figure}[t]
\centering
\begin{tikzpicture}[scale=1.05]
\fill[gray!12] (0,0) ellipse (0.42 and 0.33); \draw[setb] (0,0) ellipse (0.42 and 0.33);
\fill[orange!25] (0,-1.7) ellipse (1.35 and 0.45); \draw[setb] (0,-1.7) ellipse (1.35 and 0.45);
\fill[gray!12] (-2.7,-3.3) ellipse (0.42 and 0.33); \draw[setb] (-2.7,-3.3) ellipse (0.42 and 0.33);
\fill[white] (0.45,-3.3) ellipse (0.42 and 0.33); \draw[setb] (0.45,-3.3) ellipse (0.42 and 0.33);
\fill[orange!25] (0.45,-4.9) ellipse (1.35 and 0.45); \draw[setb] (0.45,-4.9) ellipse (1.35 and 0.45);
\fill[gray!12] (-1.4,-6.5) ellipse (0.42 and 0.33); \draw[setb] (-1.4,-6.5) ellipse (0.42 and 0.33);
\fill[gray!12] (2.3,-6.5) ellipse (0.42 and 0.33); \draw[setb] (2.3,-6.5) ellipse (0.42 and 0.33);
\fill[gray!12] (4.45,-3.3) ellipse (0.33 and 0.33); \draw[setb] (4.45,-3.3) ellipse (0.33 and 0.33);
\fill[gray!12] (5.55,-3.3) ellipse (0.33 and 0.33); \draw[setb] (5.55,-3.3) ellipse (0.33 and 0.33);
\node[act] (b1) at (0,0) {}; \node[act] (a1) at (-0.7,-1.7) {}; \node[inact] (b2) at (0.7,-1.7) {};
\node[act] (b3) at (-2.7,-3.3) {}; \node[inact] (w) at (0.45,-3.3) {};
\node[act] (a2) at (-0.25,-4.9) {}; \node[inact] (c2) at (1.15,-4.9) {};
\node[act] (c1) at (-1.4,-6.5) {}; \node[act] (c3) at (2.3,-6.5) {};
\node[act] (x) at (4.45,-3.3) {}; \node[act] (y) at (5.55,-3.3) {};
\node[nl] at (-0.62,0) {$b_1$}; \node[nl] at (-1.08,-1.7) {$a_1$}; \node[nl] at (1.08,-1.7) {$b_2$};
\node[nl] at (-3.3,-3.3) {$b_3$}; \node[nl] at (-0.22,-3.3) {$g$}; \node[nl] at (-0.63,-4.9) {$a_2$}; \node[nl] at (1.53,-4.9) {$c_2$};
\node[nl] at (-2.0,-6.5) {$c_1$}; \node[nl] at (2.9,-6.5) {$c_3$}; \node[nl] at (4.45,-3.85) {$x$}; \node[nl] at (5.55,-3.85) {$y$};
\draw[fo] (b1)--(b2); \draw[crs] (a1)--(b1);
\draw[fo] (b2)--(b3); \draw[crs] (a1)--(b3);
\draw[fo] (a1)--(w); \draw[crs] (w)--(b2);
\draw[fo] (w)--(a2); \draw[crs] (w)--(c2);
\draw[fo] (c2)--(c1); \draw[crs] (a2)--(c1);
\draw[fo] (c2)--(c3); \draw[crs] (a2)--(c3);
\draw[fo] (x)--(y); \draw[inn] (x) to[bend left=50] (y);
\node[font=\footnotesize,anchor=west] at (1.4,-1.45) {active, disconnected};
\node[font=\footnotesize,anchor=west] at (0.92,-3.3) {inactive, connected};
\node[font=\footnotesize,anchor=west] at (1.85,-4.65) {active, disconnected};
\node[font=\footnotesize,text=blue!60!black] at (-2.9,-0.3) {cluster 1};
\node[font=\footnotesize,text=red!60!black] at (5.0,-2.4) {cluster 2};
\end{tikzpicture}
\caption{The regions of the two clusters in Figure~\ref{fig:regions}, with the forest edges and the mixed edges. The twin edges $xa_1$ are not shown.
The orange regions $\{a_1,b_2\}$ and $\{a_2,c_2\}$ are disconnected, since each of them contains nodes of two components of the forest.
All other regions are connected, and $\{g\}$ is the only inactive region.}
\label{fig:clusters}
\end{figure}%
}

\newcommand{\FigureClusterTree}{%
\begin{figure}[t]
\centering
\begin{tikzpicture}[
  rbox/.style={draw=black!70,thick,rounded corners=2pt,fill=white,minimum width=0.95cm,minimum height=0.6cm,font=\small}]
\newcommand{\marc}[3]{
  \draw[fo] ($(##1)!2.6pt!90:(##2)$) -- ($(##2)!2.6pt!-90:(##1)$);
  \draw[##3] ($(##1)!2.6pt!-90:(##2)$) -- ($(##2)!2.6pt!90:(##1)$);}
\node[rbox,minimum width=1.3cm] (R) at (0,0) {};
\node[dot] (r) at (-0.12,0) {}; \node[font=\small] at (0.15,0.02) {$r$};
\node[slab,anchor=east] at (-0.75,0.03) {root region};
\node[rbox] (X) at (-2.4,-1.7) {$X$};
\node[rbox] (Y) at (0.4,-1.7) {$Y$};
\node[rbox] (Z) at (3.5,-1.7) {$Z$};
\node[rbox] (X1) at (-3.5,-3.4) {$X_1$};
\node[rbox] (X2) at (-1.3,-3.4) {$X_2$};
\node[rbox] (Y1) at (0.4,-3.4) {$Y_1$};
\node[rbox] (Z1) at (3.5,-3.4) {$Z_1$};
\begin{scope}[on background layer]
\node[fill=blue!6,draw=blue!40,rounded corners=0.6cm,inner sep=10pt,fit=(R)(X)(Y)(X1)(X2)(Y1)] (C1) {};
\node[fill=red!6,draw=red!40,rounded corners=0.5cm,inner sep=10pt,fit=(Z)(Z1)] (C2) {};
\draw[black!75,thick,dash dot,rounded corners=0.35cm] ($(X1.south west)+(-0.18,-0.2)$) rectangle ($(X2.east |- X.north)+(0.18,0.2)$);
\marc{X}{R}{crs}
\marc{Y}{R}{inn}
\marc{X1}{X}{crs}
\marc{X2}{X}{crs}
\marc{Y1}{Y}{crs}
\marc{Z1}{Z}{crs}
\draw[twn] ($(Z)!2.6pt!90:(R)$) -- ($(R)!2.6pt!-90:(Z)$);
\draw[twn] ($(Z)!2.6pt!-90:(R)$) -- ($(R)!2.6pt!90:(Z)$);
\end{scope}
\node[font=\footnotesize,text=blue!60!black,anchor=north west] at ($(C1.north west)+(0.15,-0.08)$) {cluster 1};
\node[font=\footnotesize,text=red!60!black,anchor=north] at ($(C2.south)+(0,-0.05)$) {cluster 2};
\node[font=\footnotesize,text=augcol,anchor=south west] at (1.65,-0.4) {twin arc};
\node[font=\footnotesize,anchor=north] at ($(X1.south)!0.5!(X2.south)+(0,-0.3)$) {$\witness(X)=X\cup\witness(X_1)\cup\witness(X_2)$};
\end{tikzpicture}
\caption{A schematic tree of regions.
Each box is a region, the root region contains $r$, and every other region has an arc to the region above it.
Removing the twin arcs splits the tree into the clusters (shaded).
The dash-dotted box shows that $\witness(X)$ is the union of the regions in the subtree below $X$. Here $X_1$ and $X_2$ are the children of $X$.}
\label{fig:clustertree}
\end{figure}%
}

\newcommand{\FigureDeg}{%
\begin{figure}[t]
\centering
\begin{tikzpicture}[scale=1.1]
\fill[gray!10] (0,0) ellipse (1.3 and 0.55); \draw[setb] (0,0) ellipse (1.3 and 0.55);
\fill[gray!10] (0,1.9) ellipse (1.0 and 0.45); \draw[setb] (0,1.9) ellipse (1.0 and 0.45);
\fill[gray!10] (-2.0,-1.9) ellipse (0.9 and 0.45); \draw[setb] (-2.0,-1.9) ellipse (0.9 and 0.45);
\fill[gray!10] (2.0,-1.9) ellipse (0.9 and 0.45); \draw[setb] (2.0,-1.9) ellipse (0.9 and 0.45);
\node[lab,anchor=west] at (1.36,0.12) {$X$};
\node[lab,anchor=west] at (1.08,1.95) {$\parent(X)$};
\node[lab,anchor=east] at (-2.96,-1.9) {$X_1$};
\node[lab,anchor=west] at (2.96,-1.9) {$X_2$};
\node[inact] (z1) at (-0.6,0) {}; \node[inact] (z2) at (0.6,0) {};
\node[dot] (w1) at (-0.35,1.9) {}; \node[dot] (w2) at (0.35,1.9) {};
\node[dot] (c1) at (-2.35,-1.9) {}; \node[dot] (c2) at (-1.65,-1.9) {};
\node[dot] (d1) at (1.65,-1.9) {}; \node[dot] (d2) at (2.35,-1.9) {};
\draw[fo] (z1)--(z2); \draw[fo] (z1)--(w1); \draw[aug] (z2)--(w2);
\draw[fo] (c2)--(z1); \draw[aug] (c1) to[bend left=12] (z1);
\draw[fo] (d1)--(z2); \draw[aug] (d2) to[bend right=12] (z2);
\end{tikzpicture}
\caption{Lemma~\ref{lem:deg} for a region $X$ with three mixed arcs: one to its parent region $\parent(X)$, and one from each of its children $X_1$ and $X_2$.
Each of them brings one forest edge and one mixed edge into $X$.
So $d(X)=3$ is both the number of forest edges in $\delta(X)$ and the number of endpoints of mixed edges in $X$.}
\label{fig:deg}
\end{figure}%
}

\newcommand{\FigureDigon}{%
\begin{figure}[t]
\centering
\begin{tikzpicture}[scale=1.05]
\foreach \xx in {0,2.5,5,7.5} \draw[fill=compcol,draw=compcol!65!black,line width=0.5pt] (\xx+0.1,-0.75) ellipse (0.66 and 1.28);
\foreach \i/\xx in {1/0,2/2.5,3/5,4/7.5}{
  \node[act] (x\i) at (\xx,0) {}; \node[act] (m\i) at (\xx,-1.5) {};
  \draw[fo] (x\i)--(m\i);
  \draw[bad] (x\i) to[bend left=38] (m\i);
  \node[spd,font=\small] at (\xx-0.27,-0.75) {$\beta$};
  \node[spd] at (\xx+0.27,-1.6) {$1$};
}
\foreach \i/\xx in {2/2.5,3/5,4/7.5} {\node[spd] at (\xx+0.27,0.03) {$1$}; \node[nl] at (\xx-0.8,-1.58) {$v_\i$}; \node[nl] at (\xx+0.95,0.08) {$u_\i$};}
\node[nl] at (0.1,-2.25) {$v_1$}; \node[nl] at (-0.62,0.3) {$u_1$}; \node[spd] at (-0.28,0.03) {$1$};
\node[act] (u) at (-1.9,-1.5) {}; \node[nl] at (-2.2,-1.5) {$g$}; \node[spd] at (-1.9,-1.22) {$1$};
\draw[bad] (u) to[bend left=20] (m1); \draw[bad] (u) to[bend right=20] (m1);
\draw[bad] (x1) to[bend left=22] (x2); \draw[bad] (x1) to[bend left=34] (x2);
\draw[bad] (x1) to[bend left=36] (x3); \draw[bad] (x1) to[bend left=44] (x3);
\draw[bad] (x1) to[bend left=44] (x4); \draw[bad] (x1) to[bend left=50] (x4);
\draw[opte] (m1)--(x2); \draw[opte] (m2)--(x3); \draw[opte] (m3)--(x4);
\draw[opte] (u)--(x1);
\draw[opte] (m4) .. controls (7.5,-4.1) and (-1.9,-4.1) .. (u);
\node[elab,text=badcol] at (0.5,-0.75) {$s_1$};
\node[elab,text=badcol] at (3.75,2.05) {$u_1u_i$ (two copies each)};
\node[elab,text=badcol] at (-0.9,-1.95) {$gv_1$};
\node[elab,text=optcol] at (1.37,-1.13) {$v_1u_2$};
\node[elab,text=optcol] at (-1.45,-0.5) {$gu_1$};
\node[elab,text=optcol] at (2.8,-3.72) {$v_4g$};
\end{tikzpicture}
\caption{The first example for $k=4$.
The black edges form the forest $H_0$, and the numbers are the rates at which the dual variables of the leaves, of the isolated node $g$ and of the components of $H_0$ grow.
The algorithm purchases the dark red edges, while the black and green edges form a Hamiltonian cycle.}
\label{fig:digon}
\end{figure}%
}

\newcommand{\FigureExits}{%
\begin{figure}[t]
\centering
\begin{tikzpicture}[x=0.8cm]
\begin{scope}
\fill[orange!15] (1,0) ellipse (1.45 and 0.55); \draw[setb] (1,0) ellipse (1.45 and 0.55);
\node[act] (p1) at (0,0) {}; \node[inact] (p2) at (1,0) {}; \node[act] (p3) at (2,0) {};
\draw[fo] (p1)--(p2)--(p3);
\node[pcap,text width=3.4cm] at (1.0,-0.8) {(a) \mbox{$d(S)=0$}: a whole component, with at least two leaves.};
\end{scope}
\begin{scope}[xshift=3.7cm]
\fill[orange!15] (1,0) ellipse (1.45 and 0.55); \draw[setb] (1,0) ellipse (1.45 and 0.55);
\node[act] (q1) at (0,0) {}; \node[inact] (q2) at (1,0) {}; \node[inact] (q3) at (2,0) {};
\draw[fo] (q1)--(q2)--(q3)--(2.95,0); \node[font=\small] at (3.25,0) {$\cdots$};
\node[pcap,text width=3.4cm] at (1.0,-0.8) {(b) \mbox{$d(S)=1$}: the far end of $S$ is a leaf.};
\end{scope}
\begin{scope}[xshift=8.25cm]
\fill[orange!15] (1,0) ellipse (1.45 and 0.55); \draw[setb] (1,0) ellipse (1.45 and 0.55);
\node[inact] (s1) at (0,0) {}; \node[inact] (s2) at (1,0) {}; \node[inact] (s3) at (2,0) {};
\draw[fo] (-0.95,0)--(s1)--(s2)--(s3)--(2.95,0);
\node[font=\small] at (-1.25,0) {$\cdots$}; \node[font=\small] at (3.25,0) {$\cdots$};
\node[pcap,text width=3.4cm] at (1.0,-0.8) {(c) \mbox{$d(S)=2$}: $S$ may contain no leaf.};
\end{scope}
\end{tikzpicture}
\caption{Lemma~\ref{lem:exits} for a set $S$ (shaded) such that $H[S]$ is connected.
It contains at least $2-d(S)$ leaves, where isolated nodes count twice.
The forest continues at the dots.}
\label{fig:exits}
\end{figure}%
}

\newcommand{\FigureFamilies}{%
\begin{figure}[t]
\centering
\begin{tikzpicture}[scale=1.2]
\begin{scope}[on background layer]
\node[comp,minimum width=3.7cm,minimum height=2.5cm] at (0,0.62) {};
\node[comp,minimum width=3.8cm,minimum height=2.0cm] at (-2.4,-0.95) {};
\node[comp,minimum width=3.8cm,minimum height=2.0cm] at (2.4,-0.95) {};
\node[comp,minimum width=3.0cm,minimum height=1.3cm] at (-2.9,1.9) {};
\end{scope}
\node[spd,font=\small] at (1.6,1.5) {$\beta$};
\node[spd,font=\small] at (-4.1,-0.35) {$\beta$};
\node[spd,font=\small] at (4.1,-0.35) {$\beta$};
\node[spd,font=\small] at (-4.5,2.5) {$\beta$};
\draw[orange!85!black,line width=0.9pt] (-1.4,-0.8) circle (0.2);
\draw[green!45!black,line width=0.9pt] (-3.5,1.9) circle (0.2);
\draw[violet,line width=0.9pt] (-2.9,1.95) ellipse (1.4 and 0.72);
\node[act] (a1) at (-1.0,1.0) {}; \node[nl] at (-0.72,1.3) {$a_1$}; \node[spd] at (-0.42,0.97) {$1$};
\node[inact] (g) at (0,0.3) {};  \node[nl] at (0,0.6) {$g$};
\node[act] (a2) at (1.0,1.0) {}; \node[nl] at (0.72,1.3) {$a_2$}; \node[spd] at (0.42,0.97) {$1$};
\node[act] (b1) at (-3.4,-0.8) {}; \node[nl] at (-3.4,-1.12) {$b_1$}; \node[spd] at (-3.62,-0.62) {$1$};
\node[inact] (b2) at (-2.4,-0.8) {}; \node[nl] at (-2.4,-1.12) {$b_2$};
\node[act] (b3) at (-1.4,-0.8) {}; \node[nl] at (-1.4,-1.2) {$b_3$}; \node[spd] at (-1.08,-0.8) {$1$};
\node[act] (c1) at (3.4,-0.8) {}; \node[nl] at (3.4,-1.12) {$c_1$}; \node[spd] at (3.62,-0.62) {$1$};
\node[inact] (c2) at (2.4,-0.8) {}; \node[nl] at (2.4,-1.12) {$c_2$};
\node[act] (c3) at (1.4,-0.8) {}; \node[nl] at (1.4,-1.12) {$c_3$}; \node[spd] at (1.13,-0.8) {$1$};
\node[act] (x) at (-2.3,1.9) {}; \node[nl] at (-2.3,2.17) {$x$}; \node[spd] at (-2.03,2.13) {$1$};
\node[act] (y) at (-3.5,1.9) {}; \node[nl] at (-3.5,2.24) {$y$}; \node[spd] at (-3.88,1.9) {$1$};
\draw[fo] (a1)--(g)--(a2); \draw[fo] (b1)--(b2)--(b3); \draw[fo] (c1)--(c2)--(c3); \draw[fo] (y)--(x);
\draw[crs] (a1)--(b1); \draw[crs] (a1)--(b3); \draw[crs] (g)--(b2);
\draw[crs] (a2)--(c1); \draw[crs] (a2)--(c3); \draw[crs] (g)--(c2);
\draw[inn] (y) to[bend right=45] (x);
\draw[twn] (x) to[bend left=14] (a1); \draw[twn] (x) to[bend right=14] (a1);
\end{tikzpicture}
\caption{An example that we use again in Figures~\ref{fig:regions} and~\ref{fig:clusters}.
It has four nontrivial components, six cross edges, one inner edge (parallel to the forest edge $xy$) and two twin edges $xa_1$.
The coloured outlines show the $2$-cuts $\delta(\{b_3\})$ (orange), $\delta(\{y\})$ (green) and $\delta(\{x,y\})$ (violet).}
\label{fig:families}
\end{figure}%
}

\newcommand{\FigureFewTransit}{%
\begin{figure}[t]
\centering
\begin{tikzpicture}[yscale=0.85]
\newcommand{\arcd}[3]{
  \draw[fo] ($(##1)!2.6pt!90:(##2)$) -- ($(##2)!2.6pt!-90:(##1)$);
  \draw[##3] ($(##1)!2.6pt!-90:(##2)$) -- ($(##2)!2.6pt!90:(##1)$);}
\node[regw,fill=orange!25,font=\footnotesize] (R) at (0,0) {disconnected};
\node[regw,align=center,font=\footnotesize] (T1) at (0,-1.6) {inactive\\[-1pt]connected};
\node[regw,fill=orange!25,font=\footnotesize] (J1) at (-2.0,-3.2) {disconnected};
\node[regn,font=\footnotesize] (A1) at (2.0,-3.2) {active};
\node[regw,align=center,font=\footnotesize] (T2) at (-2.0,-4.8) {inactive\\[-1pt]connected};
\node[regn,font=\footnotesize] (A2) at (-2.0,-6.4) {active};
\begin{scope}[on background layer]
\arcd{R}{T1}{crs}
\arcd{T1}{J1}{crs}
\arcd{T1}{A1}{inn}
\arcd{J1}{T2}{crs}
\arcd{T2}{A2}{inn}
\end{scope}
\draw[->,thick,orange!80!black] (T1.west) to[bend right=55] node[left=3pt,font=\footnotesize,inner sep=1pt,pos=0.4] {takes} (J1.north east);
\draw[->,thick,orange!80!black] (T2.east) to[bend left=45] node[right=2pt,font=\footnotesize,pos=0.45] {takes} ($(T2)!0.5!(A2)+(0.15,0)$);
\node[slab,anchor=west] at (1.3,0.05) {highest region};
\end{tikzpicture}
\caption{The tokens in the proof of Lemma~\ref{lem:fewtransit}, on a part of a cluster with its highest region at the top.
Grey regions are active, white regions are inactive and connected, and orange regions are disconnected.
Every white region takes one token below it: the token of a child if the arc between them is a cross arc, and the token of this arc if it is an inner arc.
Here the highest region is disconnected, and its token is never taken.}
\label{fig:fewtransit}
\end{figure}%
}

\newcommand{\FigureHub}{%
\begin{figure}[t]
\centering
\begin{tikzpicture}
\draw[fill=compcol,draw=compcol!65!black,line width=0.5pt] (1.15,0) ellipse (2.1 and 2.5);
\draw[fill=compcol,draw=compcol!65!black,line width=0.5pt] (4.6,-1.55) ellipse (0.6 and 2.2);
\node[spd,font=\small] at (-0.95,-2.1) {$\beta$}; \node[spd,font=\small] at (5.45,-1.55) {$\beta$};
\node[inact] (z) at (0,0) {}; \node[nl] at (-0.3,0.05) {$z$};
\foreach \j/\yy/\bd/\ly in {1/1.6/left/-0.25,2/0.8/left/-0.25,3/0/left/-0.25,4/-0.8/right/0.25,5/-1.6/right/0.25}{
  \node[act] (l\j) at (2.2,\yy) {};
  \draw[fo] (z)--(l\j);
  \draw[opte] (z) to[bend \bd=13] (l\j);
  \node[spd] at (2.2,\yy+\ly) {$1$};
}
\node[nl] at (2.02,1.9) {$l_1$}; \node[nl] at (2.02,-1.92) {$l_5$};
\node[act] (p0) at (4.6,0) {}; \node[act] (p1) at (4.6,-3.1) {};
\node[nl] at (5.4,0.05) {$p_0$}; \node[nl] at (5.4,-3.1) {$p_1$};
\node[spd] at (4.85,-0.27) {$1$}; \node[spd] at (4.85,-2.85) {$1$};
\draw[fo] (p0)--(p1);
\foreach \j in {1,...,5} \draw[bad] (l\j)--(p0);
\draw[bad] (p1) .. controls (2.5,-3.9) and (-0.6,-3.2) .. (z);
\draw[opte] (p1) .. controls (2.3,-4.5) and (-1.1,-3.6) .. (z);
\draw[opte] (p0) .. controls (4.7,4.6) and (-2.8,3.4) .. (z);
\node[elab,text=optcol] at (5.0,1.6) {$p_0z$};
\node[elab,text=badcol] at (3.9,1.08) {$p_0l_j$};
\node[elab,text=badcol] at (3.5,-2.98) {$p_1z$};
\node[elab,text=optcol] at (1.3,-3.9) {$p_1z$};
\end{tikzpicture}
\caption{The second example for $K=5$, drawn as in Figure~\ref{fig:digon}.
The algorithm purchases the dark red edges, while the black and green edges form a cheaper feasible solution.}
\label{fig:hub}
\end{figure}%
}

\newcommand{\FigureNonempty}{%
\begin{figure}[t]
\centering
\begin{tikzpicture}[scale=1.1]
\draw[setb] (0,0) ellipse (2.2 and 1.2);
\fill[orange!35] (-0.9,-0.15) ellipse (0.72 and 0.5);
\draw[setb] (-0.9,-0.15) ellipse (0.72 and 0.5);
\node[lab] at (-2.45,0.85) {$W$}; \node[lab] at (-0.9,-0.88) {$W'$};
\node[slab] at (0.75,-0.5) {region $\region(W)$};
\node[dot] (y1) at (-1.05,-0.3) {}; \node[dot] (o) at (-1.05,1.95) {};
\node[dot] (y2) at (-0.55,-0.05) {}; \node[dot] (x1) at (0.85,0.15) {};
\draw[aug] (y1)--node[elab,left,pos=0.8] {$e$} (o);
\draw[aug] (y2)--node[elab,above] {$\hat e_{W'}$} (x1);
\node[slab,anchor=west] at (-0.9,1.95) {outside $W$};
\end{tikzpicture}
\caption{The proof of Lemma~\ref{lem:nonempty}.
If the endpoint of $e$ in $W$ lay in a child $W'$ of $W$, then $\delta(W')$ would consist of $e$ and a primary edge $\hat e_{W'}$ of $W'$.
Both are augmentation edges, so $W'$ would be twin, which contradicts Lemma~\ref{lem:witshape}(c).}
\label{fig:nonempty}
\end{figure}%
}

\newcommand{\FigureRegions}{%
\begin{figure}[t]
\centering
\begin{tikzpicture}[yscale=0.95]
\node[regn] (R0) at (0,0) {$b_1$};
\node[regn] (R1) at (0,-1.5) {$a_1,b_2$};
\node[regn] (R2) at (-3.2,-3.0) {$b_3$};
\node[regw] (R3) at (0,-3.0) {$g$};
\node[regn] (R4) at (3.6,-3.0) {$x$};
\node[regn] (R5) at (0,-4.5) {$a_2,c_2$};
\node[regn] (R6) at (-1.7,-6.0) {$c_1$};
\node[regn] (R7) at (1.7,-6.0) {$c_3$};
\node[regn] (R8) at (3.6,-4.5) {$y$};
\begin{scope}[on background layer]
\node[fill=blue!6,draw=blue!40,rounded corners=0.9cm,inner sep=9pt,fit=(R0)(R2)(R5)(R6)(R7)] {};
\node[fill=red!6,draw=red!40,rounded corners=0.6cm,inner sep=8pt,fit=(R4)(R8)] {};
\end{scope}
\newcommand{\darc}[4]{
  \draw[fo] ($(##1)!2.6pt!90:(##2)$) -- ($(##2)!2.6pt!-90:(##1)$);
  \draw[##3] ($(##1)!2.6pt!-90:(##2)$) -- node[elab,auto=right,inner sep=3pt] {##4} ($(##2)!2.6pt!90:(##1)$);}
\begin{scope}[on background layer]
\darc{R1}{R0}{crs}{$a_1b_1$}
\darc{R2}{R1}{crs}{$a_1b_3$}
\darc{R3}{R1}{crs}{$gb_2$}
\darc{R5}{R3}{crs}{$gc_2$}
\darc{R6}{R5}{crs}{$a_2c_1$}
\darc{R7}{R5}{crs}{$a_2c_3$}
\darc{R8}{R4}{inn}{}
\node[elab,anchor=west] at (4.55,-3.75) {inner edge parallel to $xy$};
\draw[twn] ($(R4)!2.6pt!90:(R1)$) -- ($(R1)!2.6pt!-90:(R4)$);
\draw[twn] ($(R4)!2.6pt!-90:(R1)$) -- ($(R1)!2.6pt!90:(R4)$);
\end{scope}
\node[elab,text=augcol,anchor=south west] at ($(R1)!0.5!(R4)+(0.05,0.12)$) {$xa_1,\ xa_1$};
\node[font=\footnotesize,text=blue!60!black,anchor=west] at (-3.75,0.2) {cluster 1};
\node[font=\footnotesize,text=red!60!black,anchor=west] at (2.95,-5.5) {cluster 2};
\end{tikzpicture}
\caption{The tree of regions of the example in Figure~\ref{fig:families}, with $r=b_1$.
Each region is drawn with its nodes, and each arc with the edges it carries. The label of an arc names its augmentation edges.
For instance, $\{g\}$ is the region of $W=\{g,a_2,c_1,c_2,c_3\}$, and its arc to $\{a_1,b_2\}$ carries $f_W=ga_1$ and $e_W=gb_2$.
The region $\{g\}$ is the only inactive region, and removing the twin arc splits the tree into the two clusters.}
\label{fig:regions}
\end{figure}%
}

\newcommand{\FigureTransit}{%
\begin{figure}[t]
\centering
\begin{tikzpicture}
\fill[gray!8] (1.5,0.75) ellipse (1.6 and 0.9); \draw[setb] (1.5,0.75) ellipse (1.6 and 0.9);
\node[lab] at (1.5,1.9) {connected region};
\node[dot] (u1) at (0.6,0.6) {}; \node[dot] (u2) at (1.6,0.6) {}; \node[dot] (u3) at (2.6,0.6) {};
\draw[fo] (u1)--(u2)--(u3);
\fill[orange!20] (7.8,0.75) ellipse (2.7 and 1.0); \draw[setb] (7.8,0.75) ellipse (2.7 and 1.0);
\node[lab] at (7.8,2.0) {the other region};
\node[dot] (v1) at (6.5,0.6) {}; \node[dot] (v2) at (9.2,0.6) {};
\draw[fo] (u3) -- node[elab,above,pos=0.4] {$f_W$} (v1);
\draw[crs] (u2) to[bend right=30] node[elab,below=2pt,pos=0.5] {$e_W$} (v2);
\draw[fo] (u1) -- (-0.6,0.6); \node[font=\small] at (-0.85,0.6) {$\cdots$};
\node[slab,align=center] at (6.5,1.15) {same\\component};
\node[slab,align=center] at (9.2,1.15) {other\\component};
\end{tikzpicture}
\caption{The proof of Lemma~\ref{lem:transit}.
The forest edge $f_W$ and the cross edge $e_W$ of a cross arc both join its two regions.
If one of the regions is connected, then $f_W$ joins it to a node of the same component of the forest, and $e_W$ joins it to a node of another component.
So the other region is disconnected.}
\label{fig:transit}
\end{figure}%
}

\newcommand{\FigureTypes}{%
\begin{figure}[t]
\centering
\begin{tikzpicture}[scale=1.25]
\begin{scope}[on background layer]
\node[comp,minimum width=5.3cm,minimum height=3.4cm] at (0,-0.55) {};
\node[comp,minimum width=3.0cm,minimum height=1.5cm] at (2.65,2.55) {};
\end{scope}
\node[spd,font=\small] at (2.15,-1.35) {$\beta$};
\node[spd,font=\small] at (4.12,2.62) {$\beta$};
\draw[orange!85!black,line width=0.9pt] (-1.6,0) circle (0.3);
\draw[violet,line width=0.9pt] (2.65,2.55) ellipse (1.32 and 0.75);
\draw[green!45!black,line width=0.9pt] (3.3,2.55) circle (0.22);
\node[act] (z) at (0,1.6) {}; \node[nl] at (-0.05,1.92) {$z$}; \node[spd] at (-0.33,1.6) {$1$};
\node[act] (a) at (-1.6,0) {}; \node[nl] at (-1.6,0.5) {$a$}; \node[spd] at (-1.6,-0.47) {$1$};
\node[act] (c) at (0,0) {}; \node[nl] at (0.27,-0.2) {$c$}; \node[spd] at (-0.22,-0.03) {$1$};
\node[act] (d) at (1.6,0) {}; \node[nl] at (1.6,-0.32) {$d$}; \node[spd] at (1.86,-0.05) {$1$};
\node[inact] (b) at (0,-1.4) {}; \node[nl] at (0,-1.7) {$b$};
\node[act] (x) at (2.0,2.55) {}; \node[nl] at (1.85,2.85) {$x$}; \node[spd] at (2.15,2.82) {$1$};
\node[act] (y) at (3.3,2.55) {}; \node[nl] at (3.3,2.92) {$y$}; \node[spd] at (3.66,2.52) {$1$};
\draw[fo] (b)--(a); \draw[fo] (b)--(c); \draw[fo] (b)--(d); \draw[fo] (x)--(y);
\draw[crs] (z)--(a); \draw[crs] (z)--(c); \draw[crs] (z)--(d);
\draw[inn] (x) to[bend right=50] (y);
\draw[twn] (z) to[bend left=16] (x); \draw[twn] (z) to[bend right=16] (x);
\begin{scope}[shift={(4.75,1.3)}]
\draw[fo] (0,0)--(0.6,0); \node[nl,anchor=west] at (0.7,0) {forest edge};
\draw[crs] (0,-0.38)--(0.6,-0.38); \node[nl,anchor=west] at (0.7,-0.38) {cross edge};
\draw[inn] (0,-0.76)--(0.6,-0.76); \node[nl,anchor=west] at (0.7,-0.76) {inner edge};
\draw[twn] (0,-1.14)--(0.6,-1.14); \node[nl,anchor=west] at (0.7,-1.14) {twin edge};
\node[act] at (0.3,-1.52) {}; \node[nl,anchor=west] at (0.7,-1.52) {active node};
\node[inact] at (0.3,-1.9) {}; \node[nl,anchor=west] at (0.7,-1.9) {inactive node};
\node[comp,minimum width=0.75cm,minimum height=0.42cm] at (0.3,-2.3) {}; \node[nl,anchor=west] at (0.7,-2.3) {component};
\end{scope}
\end{tikzpicture}
\caption{The three kinds of augmentation edges.
The cross edge $za$ and the forest edge $ab$ form $\delta(\{a\})$, so $\{a\}$ (orange) is a mixed witness of $za$, and the same holds for $zc$ and $zd$.
The inner edge is parallel to the forest edge $xy$, and the two edges form $\delta(\{y\})$, so $\{y\}$ (green) is a mixed witness of the inner edge.
The two twin edges $zx$ form $\delta(\{x,y\})$, so $\{x,y\}$ (violet) is a twin witness of both.}
\label{fig:types}
\end{figure}%
}

\newcommand{\FigureWitness}{%
\begin{figure}[t]
\centering
\begin{tikzpicture}
\begin{scope}
\fill[red!10] (0.8,0) ellipse (1.35 and 0.48);
\draw[red!70!black,thick] (0.8,0) ellipse (1.35 and 0.48);
\draw[black!45,thick,dashed] (0,0) circle (0.27);
\draw[black!45,thick,dashed] (1.6,0) circle (0.27);
\node[act] (p) at (0,0) {}; \node[act] (q) at (1.6,0) {}; \node[act] (z) at (0.8,1.45) {};
\node[nl] at (-0.05,-0.68) {$p$}; \node[nl] at (1.65,-0.68) {$q$}; \node[nl] at (0.8,1.75) {$z=r$};
\draw[fo] (p)--(q);
\draw[twn] (p)--(z); \draw[twn] (q)--(z);
\node[pcap,text width=3.6cm] at (0.8,-1.0) {(a) The twin edges $pz$ and $qz$ have the witness set $\{p,q\}$.};
\end{scope}
\begin{scope}[xshift=4.7cm]
\fill[red!10] (-1.0,0) circle (0.27); \draw[red!70!black,thick] (-1.0,0) circle (0.27);
\node[act] (z) at (0,1.45) {}; \node[nl] at (0,1.75) {$z=r$};
\node[act] (a) at (-1.0,0) {}; \node[nl] at (-1.45,0.05) {$a$};
\node[act] (c) at (0,0) {}; \node[nl] at (0.24,-0.12) {$c$};
\node[act] (d) at (1.0,0) {}; \node[nl] at (1.25,0.08) {$d$};
\node[inact] (b) at (0,-0.8) {}; \node[nl] at (0.3,-0.86) {$b$};
\draw[fo] (b)--(a); \draw[fo] (b)--(c); \draw[fo] (b)--(d);
\draw[crs] (z)--(a); \draw[crs] (z)--(c); \draw[crs] (z)--(d);
\node[pcap,text width=3.6cm] at (0,-1.2) {(b) The cross edge $za$ has the witness set $\{a\}$.};
\end{scope}
\begin{scope}[xshift=8.7cm]
\draw[black!45,thick,dashed] (0.65,-0.06) ellipse (1.05 and 0.62);
\fill[red!10] (1.2,0) circle (0.27); \draw[red!70!black,thick] (1.2,0) circle (0.27);
\node[act] (u) at (-0.5,1.3) {}; \node[nl] at (-0.5,1.6) {$u=r$};
\node[inact] (v) at (0.15,0) {}; \node[nl] at (0.15,-0.26) {$v$};
\node[act] (w) at (1.2,0) {}; \node[nl] at (1.88,0) {$g$};
\draw[fo] (u)--(v)--(w);
\draw[inn] (u) to[bend left=25] (w);
\node[pcap,text width=3.6cm] at (0.4,-1.0) {(c) The inner edge $ug$ has the witness set $\{g\}$.};
\end{scope}
\end{tikzpicture}
\caption{Witness sets in three small examples.
The witness set is shaded red, and a dashed outline marks another witness of the edge that avoids $r$.
(a) The sets $\{p\}$ and $\{q\}$ are mixed witnesses, so the rule does not choose them for the twin edges.
Without the rule, both would be witness sets, and the forest edge $pq$ would lie in both of their cuts.
(b) The only witness of $za$ that avoids $r$ is $\{a\}$.
(c) Both $\{g\}$ and $\{v,g\}$ witness $ug$ and avoid $r$, and the witness set is the smaller one.}
\label{fig:witness}
\end{figure}%
}

\title{Better-than-3 Approximation for \twoecss \texorpdfstring{\\}{} Using the Primal-Dual Method}

\author{
Ana Cezara Danciu\thanks{
École Polytechnique de Paris.
Email: \href{mailto:ana-cezara.danciu@polytechnique.edu}%
{ana-cezara.danciu@polytechnique.edu}.  
This work was done when the author was at the Department of Computer Science, ETH Zurich, Switzerland. 
}
\and
Sharat Ibrahimpur\thanks{
Department of Computer Science, ETH Zurich, Switzerland.
Email: \href{mailto:sharat.ibrahimpur@inf.ethz.ch}%
{sharat.ibrahimpur@inf.ethz.ch}.
}
}

\begin{document}

\maketitle

\begin{abstract}
In the \emph{$2$-edge-connected spanning subgraph problem} (\twoecss), we are given an undirected multigraph whose edges have nonnegative costs, and the goal is to find a $2$-edge-connected spanning subgraph with minimum total cost. 
Two distinct approaches yield $2$-approximation algorithms: a reduction to the minimum-cost rooted $2$-arborescence problem and iterative rounding. 
Despite significant efforts, the best known approximation ratio using the primal-dual method has remained at $3$. 

\smallskip

Our main result in this work is a primal-dual algorithm for \twoecss that achieves an approximation ratio of $\sqrt{3}+1 (\leq 2.74)$. 
The algorithm runs in a single phase and the approximation guarantee is relative to the natural linear relaxation, which requires the sum of the edge variables across every cut to be at least $2$. 
The algorithm follows the standard template of growing dual variables corresponding to minimal violated sets, albeit at non-uniform rates, until the dual constraint for an edge becomes tight, at which point the edge is temporarily purchased. 
Once feasibility is attained, redundant edges are dropped in the reverse order in which they were purchased. 

\smallskip

Our choice of dual growth rates is inspired by the single-phase $3$-approximation algorithm of Beyer, Chimani, and Spoerhase~(COCOON 2020). 
As in their work, our algorithm maintains a forest of purchased edges by repeatedly contracting $2$-edge-connected subgraphs, and dual variables corresponding to leaves and isolated vertices of this forest are grown at a rate of $1$. 
Our main modification is to additionally grow dual variables corresponding to nontrivial connected components at rate $(\sqrt{3}-1)/2 \approx 0.37$, when the forest is disconnected.  
\end{abstract}


\section{Introduction}\label{sec:intro}

In this work, we consider the minimum-cost $2$-edge-connected spanning subgraph problem (\twoecss): given an undirected multigraph $G = (V,E)$ with nonnegative costs $c \colon E \to \Rp$ on its edges, find a subset of edges $F \sse E$ such that the spanning subgraph $(V,F)$ is $2$-edge-connected and the total cost $c(F) = \sum_{e \in F} c_e$ is minimized. 
In particular, each edge of $G$ may be selected at most once; parallel edges are treated as distinct edges. 
Recall that $(V,F)$ is $2$-edge-connected if for any two distinct vertices $u,v \in V$, there exist two edge-disjoint $uv$-paths using edges only from $F$. 
Whereas the $1$-edge-connected spanning subgraph problem reduces to finding a minimum-cost spanning tree and is therefore solvable in polynomial time, \twoecss is $\NP$-hard~\cite{GareyJ79}.
In fact, the problem is $\APX$-hard even with unit edge costs~\cite{CsabaKK02}, so assuming $\Pclass \neq \NP$, a polynomial-time approximation scheme (PTAS) is ruled out. 

The Generalized Steiner Network Problem (GSNP) is a central problem in survivable network design, and \twoecss is one of its simplest nontrivial special cases. 
In an instance of GSNP, we are given nonnegative integer connectivity requirements $r_{uv}$ for all pairs of distinct vertices $u,v \in V$, and we seek a minimum-cost subset of edges $F \sse E$ such that $(V,F)$ contains at least $r_{uv}$ edge-disjoint $uv$-paths for every pair $u,v$.
In this formulation, \twoecss corresponds to setting $r_{uv} = 2$ for all distinct $u,v \in V$.
Thus, \twoecss captures the basic fault-tolerance requirement that the network remain connected after any single edge failure.

Because survivable network design problems arise in numerous real-world applications, substantial attention has been devoted to developing approximation algorithms for these problems. 
One of the earliest results for \twoecss is due to Frederickson and J\'aJ\'a~\cite{FredericksonJ81}, who gave a $3$-approximation algorithm by augmenting a minimum-cost spanning tree.
Khuller and Vishkin~\cite{KhullerV94} subsequently improved the approximation ratio to $2$ via a reduction to finding a minimum-cost pair of arc-disjoint rooted spanning arborescences in the bidirected version of the input graph $G$.
Their approach also extends to the more general $\kecss$ problem, where a fixed integer $k\geq 2$ is prescribed and $r_{uv}=k$ for all distinct $u,v\in V$, by computing a minimum-cost collection of $k$ arc-disjoint rooted spanning arborescences.

A different line of work approaches survivable network design using the primal-dual method.
Agrawal, Klein and Ravi~\cite{AgrawalKR95} initiated this line of work, obtaining a $2\lceil\log_2(k+1)\rceil$-approximation for GSNP when $r_{uv}\leq k$ for all distinct $u,v\in V$.
Their model allows an edge $e$ to be purchased multiple times, with each copy incurring a cost of $c_e$.
Williamson, Goemans, Mihail and Vazirani~\cite{WilliamsonGMV95} subsequently considered the setting in which each edge can be purchased at most once, and obtained a $2H_k$-approximation for $\kecss$, where $H_k=\sum_{i=1}^k 1/i$ is the $k$-th harmonic number.
Thus, for \twoecss, their approximation ratio is $3$. 
Their algorithm proceeds in $k$ phases, increasing the connectivity requirement by one in each phase.
Within each phase, it repeatedly grows the dual variables corresponding to the inclusion-minimal violated sets until edges become tight and are purchased, and then performs a reverse-delete step to remove redundant purchased edges.

A major breakthrough was made by Jain~\cite{Jain01}, who introduced the technique of \emph{iterative rounding} and gave a $2$-approximation for GSNP with arbitrary pairwise connectivity requirements.
Jain's algorithm is based on the natural cut linear programming (LP) relaxation and a key property of its extreme point solutions: every extreme point solution to the residual LP contains an edge of value at least $1/2$.
Purchasing such an edge and iterating yields the $2$-approximation guarantee.

Thus, two quite different approaches---rooted arborescences and iterative rounding---give a $2$-approximation for \twoecss.
It is therefore intriguing that, despite more than three decades of active research on primal-dual methods for network design, the best approximation ratio obtained by this framework for \twoecss has remained at $3$.
More recently, Beyer, Chimani and Spoerhase~\cite{BeyerCS20} gave an insightful single-phase primal-dual algorithm that constructs a $2$-edge-connected solution directly, rather than first constructing a spanning tree and then augmenting it to become $2$-edge-connected.
Their approximation ratio is also $3$, and this guarantee is tight.

\subsection{Our contributions}\label{sec:results}

Our main contribution is the first primal-dual approximation algorithm for \twoecss with an approximation ratio strictly better than $3$.

\begin{theorem} \label{thm:main}
There is a single-phase primal-dual algorithm for \twoecss that achieves an approximation ratio of $\sqrt{3}+1$ with respect to the natural cut LP relaxation. 
The running time of the algorithm is polynomial in the input size. 
\end{theorem}

Our algorithm builds on the single-phase algorithm of~\cite{BeyerCS20}.
At any point in the algorithm, contracting the $2$-edge-connected subgraphs of the currently purchased solution $(V,F)$ results in a forest. 
The leaves, isolated nodes, and nontrivial connected components of this forest correspond to minimal violated sets of different deficiencies.
As in~\cite{BeyerCS20}, we grow the dual variables corresponding to leaves and isolated nodes at rate $1$.
Our main algorithmic change is to additionally grow the dual variables corresponding to nontrivial connected components at a rate $\beta\in[0,1/2]$, when the forest is disconnected. 
Setting $\beta=0$ recovers their algorithm, whereas our approximation guarantee in \Cref{thm:main} is obtained by choosing $\beta = (\sqrt{3}-1)/2$. 

Our analysis follows the standard primal-dual template of bounding incidences between active sets and edges that remain after reverse delete.
The new feature is that our dual variables are grown at nonuniform rates, so these incidences must be counted with corresponding weights.
After reverse delete, the surviving edges $A$ form a minimal augmentation of the current forest, that is, $(V,F \sqcup A)$ is $2$-edge-connected but $(V,F\sqcup(A\sm\{e\}))$ is not $2$-edge-connected for any $e\in A$.
Accordingly, the weight (or implied cost) of an augmentation edge is the number of its incidences with leaves and isolated nodes, plus $\beta$ times the number of nontrivial connected components it crosses.
The crux of our analysis is the following combinatorial bound.

\begin{theorem}\label{thm:aug}
Let $H=(V,F)$ be a forest and $A$ be a minimal augmentation of $H$. 
Let $L$ and $I$ denote the sets of leaf vertices (degree $1$) and isolated vertices (degree $0$) of $H$, let $\C$ denote the collection of vertex sets of its nontrivial connected components, and for $S \sse V$, let $a(S)$ be the number of edges of $A$ with exactly one endpoint in $S$.
Then, for any $\beta \in [0,1/2]$, we have: 
\begin{equation}\label[ineq]{eq:aug}
\sum_{v \in L \cup I} a(v) + \beta \sum_{C \in \C} a(C) \leq
2(1+\beta) \cdot \bigl( |L| + 2|I| + \frac{|\C|}{1+\beta}  - 2 \bigr).
\end{equation}
\end{theorem}

We use \Cref{thm:aug} to show that, for any $\beta \in [0,1/2]$, our algorithm pays at most
${\max\{2+2\beta,\,2+\frac{1}{1+\beta}\}}$ times what the dual earns.
This ratio is minimized at $\beta = \frac{\sqrt3-1}{2}$, where it equals $\sqrt3+1$, yielding \Cref{thm:main}. 
 
\subsection{Related work}\label{subsec:relatedwork}

No approximation ratio below $2$ is known for \twoecss with general edge costs, but better guarantees are known in several special cases.
For unit edge costs, a long line of work~\cite{CheriyanSS01,SeboV14,HunkenschroderVV19,GargGJ23,KobayashiN23,BoschCalvoGGHJL25} has culminated in an approximation ratio strictly below $5/4$~\cite{HommelsheimLL26}.
See~\cite{BoydIT13,Takazawa16} for further specializations. 
There is also an extensive literature on restricted costs, in particular when edge costs lie in $\{0,1\}$ and/or when the zero-cost edges are required to have a prescribed structure, such as a forest, a tree, a collection of paths, or a matching. 

The natural cut LP relaxation for \twoecss has integrality gap between $3/2$~\cite{CheriyanKKK08} and $2$~\cite{Jain01}.
For unit edge costs, the gap lies between $8/7$~\cite{BoydFS16} and $4/3$~\cite{SeboV14}.
For series-parallel graphs, the relaxation is integral~\cite{Mahjoub94}.

For planar graphs, PTASes are known for unit costs~\cite{CzumajGSZ04} and general costs~\cite{BergerG07}.
Exact polynomial-time algorithms are known for series-parallel graphs~\cite{Winter86} and, more generally, graphs of bounded treewidth~\cite{BergerG07}.

\twoecss has also been studied in a model where an edge $e$ may be purchased multiple times, with each copy incurring cost $c_e$. 
This model is more permissive than ours, and approximation guarantees for it do not directly apply to our setting.
For instance, if the input graph $G$ has a bridge, then the instance is infeasible in our setting, whereas a solution in the multisubgraph model can buy the bridge twice. 

\medskip

\noindent 
\textbf{Organization of the paper.}
\Cref{sec:notation} introduces our notation and the cut LP relaxation, \Cref{sec:algo} presents our algorithm, and \Cref{sec:minaug} gives a proof of \Cref{thm:aug}.

\section{Preliminaries}\label{sec:notation}

We reserve the notation $(G=(V,E),c \colon E \to \Rp)$ for an instance of $\twoecss$, where $G$ is a multigraph on $n \coloneqq |V|$ vertices and $c$ is a nonnegative cost-function on the edges. 
The multigraphs that we encounter may have parallel edges but no (self) loops. 
An edge-set $F \sse E$ is \emph{feasible} for the given \twoecss instance if the spanning subgraph $(V,F)$ is $2$-edge-connected. 
We shall assume that $E$ is feasible, since otherwise the instance is infeasible, and that $|E| \leq n(n-1)$, since any inclusion-minimal feasible solution never contains more than two edges with the same pair of endpoints.
We use the shorthand $c(F) \coloneqq \sum_{e \in F} c_e$ to denote the total cost of edges in $F$, and use $\OPT$ to denote the cost of an optimal solution. 

An algorithm for \twoecss is said to be \emph{efficient} if its running time is polynomial in $n$ and the binary encoding length of the edge costs.  

For a multigraph $H$ and $\emptyset \neq S \ssne V(H)$, we use $\del_H(S) \coloneqq \{ e \in E(H) : |S \cap V(e)| = 1 \}$ to denote the set of edges with exactly one endpoint in $S$. 
We omit the subscript when it is clear from the context and use $\del(v) = \del(\{v\})$ to avoid cumbersome notation. 
We use $\deg(v) = |\del(v)|$ to denote the degree of $v$ and $\kappa(H)$ to denote the number of connected components of $H$, including isolated vertices.
When $H$ is connected, we refer to $S$ and $\del(S)$ interchangeably as a cut of $H$.

We now describe a natural linear relaxation for \twoecss, referred to as the cut LP. 
For each edge $e \in E$, the (primal) LP has a nonnegative variable $x_e$ that models the portion of edge $e$ that is purchased in an optimal solution. 
\Cref{primalcovering} models the requirement that at least two units of edges are purchased from any cut of $G$, while \Cref{primalupperbound} models the requirement that no edge is purchased more than once.  
The corresponding dual LP has nonnegative variables $\{\yvar{S}\}_{\emptyset \neq S \ssne V}$ and $\{z_e\}_{e \in E}$ corresponding to the primal cut constraints and upper-bound constraints, respectively.
For each edge $e \in E$, \Cref{dualpacking} requires that the total $y$-value of cuts containing $e$, minus $z_e$, is at most $c_e$.
For any fixed choice of the $y$-variables, the objective is maximized by taking
$z_e = \max\{0,\sum_{S \colon e \in \delta(S)} \yvar{S} - c_e\}$.
In particular, once \cref{dualpacking} becomes tight for an edge $e$, any further increase in the $y$-variables appearing on its left-hand side is accompanied by an equal increase in $z_e$. Clearly, the cut LP is a valid relaxation of \twoecss, and hence its optimum value is at most $\OPT$.
Moreover, by LP weak duality, the objective value of any feasible dual solution (as given by \eqref{eq:dualobj}) is at most the primal LP optimum, and therefore also at most $\OPT$. 

\medskip

\noindent
\begin{minipage}[t]{0.485\textwidth}
\begin{center}
    \textbf{Primal LP}
\end{center}
\vspace{-10pt}
\begin{subequations}\label{eq:primalLP}
\begin{alignat}{2}
\min \ \ && \sum_{e \in E} c_e x_e \label{eq:primalobj} \\
\text{s.t.} \ \ && \sum_{e \in \delta_G(S)} x_e & \geq 2 \quad \forall\, \emptyset \neq S \ssne V \label[cons]{primalcovering} \\[9pt]
&& x_e & \leq 1 \quad \forall\, e \in E \label[cons]{primalupperbound} \\
&& x_e & \geq 0 \quad \forall\, e \in E \label[cons]{primalnonneg}
\end{alignat}
\end{subequations}
\end{minipage}%
\hspace{0.25em}\vline\hspace{-0.1em}%
\begin{minipage}[t]{0.485\textwidth}
\begin{center}
    \textbf{Dual LP}
\end{center}
\vspace{-10pt}
\begin{subequations}\label{eq:dualLP}
\begin{alignat}{1}
    \max & \sum_{\emptyset \neq S \ssne V} 2 \yvar{S}
    - \sum_{e \in E} z_e \label{eq:dualobj} \\
    \text{s.t.} & \sum_{\substack{\emptyset \neq S \ssne V \colon \\ e \in \delta_G(S)}} \yvar{S} - z_e \leq c_e \quad \forall\, e \in E \label[cons]{dualpacking} \\
    & \qquad \qquad z_e \geq 0 \qquad \quad \ \forall\, e \in E
    \label[cons]{dualznonneg} \\
    & \qquad \qquad \yvar{S} \geq 0 \quad \forall\, \emptyset \neq S \ssne V \label[cons]{dualynonneg}
\end{alignat}
\end{subequations}
\end{minipage}

\smallskip

\section{Our Algorithm} \label{sec:algo}

In this section, we describe a primal-dual algorithm parameterized by $\beta \in [0,1/2]$ that outputs a feasible solution $F \sse E$ and a dual feasible solution $(y,z)$. 
In \Cref{lem:guarantee}, we show that $c(F)$ is at most $\max\{2+2\beta,2+\frac{1}{1+\beta}\}$ times the dual objective value of $(y,z)$ (as given by \eqref{eq:dualobj}). 
The proof of \Cref{thm:main} will follow from running the algorithm with $\beta = (\sqrt{3}-1)/2$. 

At a high level, our algorithm maintains a contracted graph $H = (V_H,E_H)$ whose nodes correspond to a partition of the vertices in $G$.
Whenever a purchased edge creates a cycle in $H$, we contract all nodes of the cycle into a single node, thereby maintaining the invariant that $H$ is a forest.
We simulate the growth of variables in the dual solution by decreasing the reduced costs of unpurchased edges in $E$ whose endpoints belong to distinct nodes of $H$.
Whenever the reduced cost of some edge $e$ reaches zero, we purchase that edge and update $H$ accordingly.
The dual growth procedure terminates when $H$ has a single node.
Finally, we perform a reverse-delete step on the purchased edges to remove redundancies while preserving feasibility.

\smallskip

\noindent The rest of the section is divided into three subsections:
\begin{itemize}[leftmargin=*, nosep]
    \item In \Cref{subsec:algodesc}, we describe the algorithm formally.
    Its dual growth raises the dual variables of the leaves and isolated nodes of $H$ at rate $1$ and, while $H$ is disconnected, those of its nontrivial components at rate $\beta$.

    \item In \Cref{subsec:analysis}, we show that the algorithm is efficient and outputs feasible primal and dual solutions, and we bound the cost of its output using \Cref{thm:aug}.
    At the start of every iteration, the edges of the output that are not yet purchased form a minimal augmentation of the current forest (\Cref{lem:minaug-iteration}).
    The total rate at which these edges are paid for is exactly the left side of \eqref{eq:aug} for this forest and augmentation, while the dual objective grows at rate $|L|+2|I|+2\beta|\C|$ when the forest is disconnected, with $L$, $I$ and $\C$ as in \Cref{thm:aug} (\Cref{lem:dualrate}).
    So \Cref{thm:aug} bounds the total rate at which these edges are paid for by $2+2\beta$ times $|L|+2|I|$, plus $2$ for every nontrivial component.
    A nontrivial component adds only $2\beta$ to the dual rate, but it has at least two leaves, and balancing these two terms leads to an approximation ratio of $\max\{2+2\beta,2+\frac{1}{1+\beta}\}$ (\Cref{lem:guarantee}). 

    \item In \Cref{subsec:tight}, we show that this analysis is tight for every $\beta\in[0,1/2]$ (\Cref{thm:tight}).
    We give two families of instances on which the ratio of the algorithm tends to $2+\frac{1}{1+\beta}$ and to $2+2\beta$, respectively.
    In particular, no choice of $\beta\in[0,1/2]$ gives our algorithm an approximation ratio below $\sqrt3+1$.
\end{itemize}

\subsection{Description of the algorithm}\label{subsec:algodesc}

We now formally describe the algorithm.
A node $u \in V_H$ stands for the set $V_u \sse V$ of vertices of $G$ contracted into it, and $V_X \coloneqq \bigcup_{u \in X} V_u$ for $X \sse V_H$.
An edge of $G$ \emph{joins} the nodes $u,v \in V_H$ if it has one endpoint in $V_u$ and the other in $V_v$, and it is \emph{relevant} if it is not purchased and joins two distinct nodes of $H$.
A node of $H$ is a \emph{leaf} if it has degree $1$ and \emph{isolated} if it has degree $0$, and a connected component of $H$ is \emph{nontrivial} if it has at least two nodes.

\begin{enumerate}[label=\upshape(S\arabic*),leftmargin=*,nosep]
\item \textbf{Input:} The algorithm takes a \twoecss instance $(G=(V,E), c : E \to \Rp)$ and a parameter $\beta \in [0,1/2]$.
\item \textbf{Initialization:}
Set $t \coloneqq 1$, let $F$ be an empty ordered list of purchased edges, and set $(y,z) \coloneqq (\mathbf{0},\mathbf{0})$.
Set $H \coloneqq (V,\emptyset)$, so that $V_u = \{u\}$ for every $u \in V$, and give every edge $e \in E$ the reduced cost $\rc_e \coloneqq c_e$.
\item \textbf{Dual growth:}
If $|V_H| = 1$, go to~(S5).
Let $L$ and $I$ be the sets of leaves and isolated nodes of $H$, and for every node $u \in V_H$, let $C_u \sse V_H$ be the node-set of the connected component of $H$ containing $u$.
For a relevant edge $e$ that joins $u$ and $v$, let
\[
w_e \coloneqq \bone{u \in L \cup I} + \bone{v \in L \cup I} + \beta \cdot \bone{C_u \neq C_v} \cdot \bigl( \bone{|C_u| \geq 2} + \bone{|C_v| \geq 2} \bigr),
\]
where $\bone{P}$ is $1$ if the statement $P$ holds and $0$ otherwise. Hence, $e$ gets $1$ for each of its endpoints in a node of $L \cup I$ and $\beta$ for each nontrivial component $C$ of $H$ with $e \in \delta_G(V_C)$.
Simultaneously, increase $\yvar{V_u}$ at rate $1$ for every node $u \in L \cup I$ and, if $H$ is disconnected, $\yvar{V_C}$ at rate $\beta$ for every nontrivial component $C$ of $H$.
At the same time, decrease $\rc_e$ at rate $w_e$ for every relevant edge $e$, and increase $z_e$ for every purchased edge $e$ at the same rate as the total $y$-value of cuts containing $e$, which keeps \Cref{dualpacking} tight for $e$.
Stop as soon as some relevant edge $e_t$ has $\rc_{e_t} = 0$.
\item \textbf{Purchase and contraction:}
Append $e_t$ to $F$, and let $u,v \in V_H$ be the nodes that $e_t$ joins.
If $C_u \neq C_v$, add $e_t$ to $E_H$.
Otherwise, $e_t$ and the $u$-$v$ path of $H$ form a cycle, and we contract its nodes into a single node $x$ with $V_x$ the union of their vertex-sets.
Increase $t$ by $1$ and go to~(S3).
\item \textbf{Reverse delete:}
For $j = t-1, t-2, \ldots, 1$, delete $e_j$ from $F$ if $(V,F \sm \{e_j\})$ is $2$-edge-connected.
\item \textbf{Output:}
Output the final edge-set $F$ together with the dual solution $(y,z)$ constructed during the dual-growth steps.
\end{enumerate}

\subsection{Analysis}\label{subsec:analysis}

An \emph{iteration} is one execution of (S3) and (S4). 
The algorithm purchases $e_t$ in iteration $t$.
For an iteration $t$, let $H_t$ be the forest $H$ at its beginning, let $L_t$ and $I_t$ be the sets of leaves and isolated nodes of $H_t$, let $\Delta_t \geq 0$ be the duration of its dual growth, and let $w^t_e$ denote the rate $w_e$ (of decrease in reduced cost) of a relevant edge $e$ during iteration $t$. 
We also let $A_t \sse F$ consist of the edges purchased in iteration $t$ or later that survive the reverse-delete step.

Let $T$ be the number of iterations, let $H_{T+1}$ be the final forest, which has a single node, and for $1 \leq t \leq T+1$, let $F_t \coloneqq \{e_1,\ldots,e_{t-1}\}$ be the set of edges purchased before iteration $t$.
The \emph{load} of an edge $e$ is $\ell_e \coloneqq \sum_{S \colon e \in \del(S)} \yvar{S}$, the left side of \cref{dualpacking} without $z_e$.
Nodes are only merged, so if an edge joins two distinct nodes of $H_s$, then it joins two distinct nodes of $H_t$ for every $t \leq s$.

The following three lemmas are consequences of the design of our algorithm.

\begin{lemma} \label{lem:pdfeasibility}
The algorithm is efficient, and it outputs a feasible solution $F$ for the given \twoecss instance and a feasible solution $(y,z)$ to the dual of the cut LP.
\end{lemma}

\begin{proof}
Every iteration either adds an edge to $H$, which decreases $\kappa(H)$ by $1$, or contracts at least two nodes into one, which decreases $|V_H|$.
So $|V_H| + \kappa(H)$ decreases in every iteration, from $2n$ to $2$, and $T \leq 2n-2$.
Each iteration takes polynomial time, provided that its dual growth ends, which we show below.
The reverse delete tests $2$-edge-connectivity at most $2n-2$ times, so the algorithm is efficient.

\smallskip
We next show by induction on $t$ that, for every $t \leq T+1$:
\begin{enumerate}[label=(\roman*),nosep]
\item $H_t$ is a forest, and its edges are the edges of $F_t$ that join two distinct nodes.
\item For every node $x$ of $H_t$, the edges of $F_t$ with both endpoints in $V_x$ form a $2$-edge-connected graph on $V_x$.
\end{enumerate}
Both hold for $t=1$.
Let $t \leq T$, and let $e_t$ join $u$ and $v$.
If $C_u \neq C_v$, then $H_t + e_t$ is a forest and the sets $V_x$ do not change.
Otherwise, let $P$ be the node-set of the cycle formed by $e_t$ and the $u$-$v$ path of $H_t$.
The new node $x$ has $V_x = V_P$, and the edges of $F_{t+1}$ inside $V_x$ consist of $2$-edge-connected graphs on the sets $V_{x'}$, $x' \in P$, and of the edges of the cycle, which join these sets in a cyclic order.
This graph is $2$-edge-connected: a cut of it either separates two vertices of some $V_{x'}$, and then contains two edges inside $V_{x'}$, or it is crossed twice by the cycle.
By (i), the edges of $F_{t+1}$ that join two nodes of $P$ are the edges of the cycle, as $H_t$ is a forest.
Contracting the nodes of a cycle of $H_t+e_t$ into one node gives a forest, and its edges are the other edges of $H_t + e_t$, so (i) and (ii) hold for $t+1$.

\smallskip

Next, the dual growth of every iteration $t \leq T$ ends.
The forest $H_t$ has at least two nodes, so it has a node $u$ with $\deg_{H_t}(u) \leq 1$, and $V_u$ is a nonempty proper subset of $V$.
As $E$ is feasible, $\delta_G(V_u)$ contains at least two edges, and each of them joins $u$ with another node.
By (i), at most one of them is purchased, so some relevant edge has an endpoint in $V_u$, and its rate is at least $1$.
In every iteration, the reduced costs of the relevant edges are nonnegative at the start, and they decrease only until the first of them reaches $0$. 
If some relevant edge already has reduced cost $0$, then $\Delta_t=0$.
Otherwise, the dual growth ends after time
$\Delta_t=\min\{\bar c_e/w_e^t : e\text{ relevant},\ w_e^t > 0\}$. 

\smallskip

At the end, $H_{T+1}$ has a single node $x$ with $V_x = V$, so by (ii), the set of all purchased edges is feasible.
The reverse delete deletes an edge only if the remaining edges are still feasible, so the output $F$ is feasible.

\smallskip

It remains to show that $(y,z)$ is feasible.
In iteration $t$, every grown set is a nonempty proper subset of $V$: a set $V_u$ is proper as $H_t$ has at least two nodes, and a set $V_C$ is proper as $H_t$ is then disconnected.
Every grown set is a union of sets $V_x$ with $x$ a node of $H_t$, so an edge with both endpoints in one set $V_x$ lies in the cut of no grown set.
As nodes are only merged, the load of an edge stops changing once its endpoints lie in one node.
A relevant edge $e$ that joins $u$ and $v$ lies in $\delta(V_x)$, for a node $x$, exactly when $x \in \{u,v\}$, and it lies in $\delta(V_C)$, for a component $C$, exactly when $C_u \neq C_v$ and $C \in \{C_u,C_v\}$.
So its load increases at rate $w^t_e$, the rate at which $\rc_e$ decreases.
Hence $\rc_e = c_e - \ell_e$ for every edge $e$ that is not purchased, at all times, and $\rc_e \geq 0$ gives $\ell_e \leq c_e$ and $z_e = 0$.
An edge $e$ purchased in iteration $t$ has $\ell_e = c_e$ and $z_e = 0$ at the end of iteration $t$, and afterwards $z_e$ increases exactly as fast as $\ell_e$, so $\ell_e - z_e = c_e$ and $z_e \geq 0$.
So $(y,z)$ satisfies all constraints of the dual LP.
\end{proof}

\begin{lemma} \label{lem:minaug-iteration}
For any iteration $t$, every edge of $A_t$ joins two distinct nodes of $H_t$, and, viewed as edges between these nodes, $A_t$ is an inclusion-minimal augmentation of $H_t$.
\end{lemma}

We use the following consequence of (i) and (ii).
For $t \leq T$ and every set $B$ of edges that join two distinct nodes of $H_t$,
\begin{equation}\label{eq:contract}
(V, F_t \cup B) \text{ is $2$-edge-connected} \iff H_t + B \text{ is $2$-edge-connected}.
\end{equation}
Indeed, let $\emptyset \neq S \ssne V$.
If $S$ separates two vertices of some set $V_x$, then by (ii), $\delta(S)$ contains at least two edges of $F_t$.
Otherwise, $S = V_X$ for a nonempty proper subset $X$ of the nodes of $H_t$, and by (i), the edges of $F_t \cup B$ in $\delta(V_X)$ are exactly the edges of $H_t + B$ in $\delta(X)$.
So both graphs have a cut with at most one edge, or neither of them has.

\begin{proof}
Let $e \in A_t$ be purchased in iteration $s \geq t$.
Then $e$ is relevant in iteration $s$, so it joins two distinct nodes of $H_s$, and hence of $H_t$.
The output $F$ is contained in $F_t \cup A_t$ and is feasible, so $H_t + A_t$ is $2$-edge-connected by \eqref{eq:contract}.
When the reverse delete examined $e$, the remaining edges were the edges purchased before $e$, the edge $e$, and the edges purchased after $e$ that are in the output. Without $e$, they were not feasible, since $e$ was kept.
These edges without $e$ include $F_t \cup (A_t \sm \{e\})$, because the edges of $F_t$ are purchased before $e$, and every edge of $A_t \sm \{e\}$ is purchased before $e$ or is in the output.
So $(V, F_t \cup (A_t \sm \{e\}))$ is not $2$-edge-connected, and neither is $H_t + (A_t \sm \{e\})$, by \eqref{eq:contract}.
\end{proof}

\begin{lemma}\label{lem:dualrate}
We have:
\begin{enumerate}[label=\upshape(\roman*)]
\item For any iteration $t \in [T]$, the reduced costs of the edges of $A_t$ decrease at total rate 
\[
w^t(A_t) \coloneqq \sum_{e \in A_t} w^t_e.
\] 
Hence, 
\[
c(F) = \sum_{t=1}^T \Delta_t\, w^t(A_t).
\]
\item The dual objective increases at rate 
\[
D(t) \coloneqq
\begin{cases}
|L_t| + 2|I_t| + 2\beta\bigl(\kappa(H_t)-|I_t|\bigr) & \text{if $H_t$ is disconnected,}\\
|L_t| & \text{otherwise.}
\end{cases}
\]
Hence, 
\[
\sum_{\emptyset \neq S \ssne V} 2\yvar{S} - \sum_{e \in E} z_e = \sum_{t=1}^T \Delta_t\, D(t).
\]
\end{enumerate}
\end{lemma}

\begin{proof}

By \Cref{lem:minaug-iteration}, every edge $e \in A_t$ joins two distinct nodes of $H_t$, and it is purchased in iteration $t$ or later, so it is relevant during iteration $t$ and $\rc_e$ decreases at rate $w^t_e$.

Increasing $\yvar{S}$ at rate $\sigma$ increases $\sum_S 2\yvar{S}$ at rate $2\sigma$, and $\sum_e z_e$ at rate $\sigma\, d_{F_t}(S)$, since $z_e$ increases with $\yvar{S}$ exactly for the purchased edges $e$ in $\delta(S)$.
The grown sets are unions of sets $V_x$, so by (i) in the proof of \Cref{lem:pdfeasibility}, the purchased edges in the cut of a grown set are edges of $H_t$.
So $d_{F_t}(V_u) = \deg_{H_t}(u)$, which is $1$ for a leaf and $0$ for an isolated node, and $d_{F_t}(V_C) = 0$ for every component $C$ of $H_t$.
Hence every leaf adds $1$ to the rate of the dual objective, every isolated node adds $2$, and, if $H_t$ is disconnected, every nontrivial component adds $2\beta$.
There are $\kappa(H_t)-|I_t|$ nontrivial components, as the components of $H_t$ with one node are its isolated nodes.
If $H_t$ is connected, it has no isolated node, as it has at least two nodes, and no component grows.
This gives the rate $D(t)$.

An edge $e \in F$ purchased in iteration $t_e$ is relevant in every iteration $t \leq t_e$, and its reduced cost decreases from $c_e$ to $0$ during these iterations, so $c_e = \sum_{t \leq t_e} \Delta_t\, w^t_e$.
As $e \in A_t$ exactly for $t \leq t_e$, summing over $e \in F$ gives $c(F) = \sum_t \Delta_t\, w^t(A_t)$.
The dual objective is $0$ at the start, and it changes only during the dual growth, at rate $D(t)$ for time $\Delta_t$ in iteration $t$.
\end{proof}

By \Cref{lem:minaug-iteration}, $A_t$ is an inclusion-minimal augmentation of the forest $H_t$ in every iteration $t$, so we may apply \Cref{thm:aug} to $H_t$ and $A_t$.
Recall that \Cref{thm:aug} bounds the number of endpoints of edges of $A_t$ at leaves and isolated nodes of $H_t$, plus $\beta$ times the total number of edges of $A_t$ in the cuts of the nontrivial components of $H_t$, in terms of the numbers of leaves, isolated nodes and components of $H_t$.
The left-hand side of this bound is exactly the rate $w^t(A_t)$ of \Cref{lem:dualrate}, while its right-hand side depends only on the quantities that determine the dual rate $D(t)$.
Comparing these two rates in every iteration leads to the following guarantee.

\begin{lemma}\label{lem:guarantee}
We have $c(F) \leq \max\{2+2\beta,2+\frac{1}{1+\beta}\} \bigl( \sum_S 2\yvar{S} - \sum_e z_e \bigr)$.
\end{lemma}
\begin{proof}
Let $\lambda \coloneqq \max\{2+2\beta,\,2+1/(1+\beta)\}$. 

Fix an iteration $t$, and let $L = L_t$, $I = I_t$ and $\kappa = \kappa(H_t)$.
As the components of $H_t$ with one node are its isolated nodes, $H_t$ has $\kappa-|I|$ nontrivial components.
By \Cref{lem:minaug-iteration}, $A_t$ is a minimal augmentation of the forest $H_t$, which has at least two nodes.
An edge of $A_t$ that joins $u$ and $v$ leaves the components $C_u$ and $C_v$ if $C_u \neq C_v$, and no component otherwise.
So $w^t_e$ is the number of endpoints of $e$ in nodes of $L \cup I$ plus $\beta$ times the number of nontrivial components of $H_t$ whose cut contains $e$, and $w^t(A_t)$ is the left side of the inequality in \Cref{thm:aug} for $H_t$ and $A_t$.

If $H_t$ is connected, then $I = \emptyset$, $\kappa = 1$, and no edge lies in the cut of a component, so \Cref{thm:aug} with $\beta = 0$ gives $w^t(A_t) \leq 2|L| - 2 \leq \lambda D(t)$.
Otherwise, \Cref{thm:aug} gives $w^t(A_t) \leq (2+2\beta)(|L|+2|I|) + 2(\kappa-|I|)$, and hence
\[
w^t(A_t) - \lambda D(t) \ \leq\ (2+2\beta-\lambda)\bigl(|L|+2|I|\bigr) + (2-2\beta\lambda)(\kappa-|I|) .
\]
The first term is at most $0$, as $\lambda \geq 2+2\beta$, so we are done if $2-2\beta\lambda \leq 0$.
Otherwise, we use that every nontrivial component has at least two leaves, so $2(\kappa-|I|) \leq |L| \leq |L|+2|I|$, and the right side is at most $\bigl(3+2\beta-(1+\beta)\lambda\bigr)(|L|+2|I|) \leq 0$, as $\lambda \geq 2+ 1/(1+\beta)$.
\end{proof}

\begin{proofof}{\Cref{thm:main}}
Run the algorithm with $\beta = \bstar = (\sqrt3-1)/2$.\footnote{To compute with rational numbers only, one can run the algorithm with a rational $\beta$ such that $|\beta-\bstar| \leq 2^{-\Theta(b)}$, where $b$ is the binary encoding length of the input, and get the same guarantee.}
Then $2+2\beta = 2+\frac{1}{1+\beta} = \sqrt3+1$, and the theorem follows from \Cref{lem:pdfeasibility,lem:guarantee}.
\end{proofof}

\subsection{Tightness of the analysis}\label{subsec:tight}

We conclude this section by showing that the cost analysis in \Cref{lem:guarantee} is tight for any $\beta \in [0,1/2]$. 

\begin{theorem}\label{thm:tight}
Fix $\beta \in [0,1/2]$ and $\eps > 0$. 
There is a \twoecss instance on which the algorithm run with parameter $\beta$ outputs a solution of cost at least ${\bigl(\max\{2+2\beta,2+\frac{1}{1+\beta}\}-\eps\bigr)\cdot\OPT}$. 
\end{theorem}

\begin{proof}
Let $\beta\in[0,\frac12]$ and $\eps>0$.
The first example below gives an instance on which the output costs at least $(2+\frac1{1+\beta}-\eps)\cdot\OPT$, and the second gives one on which it costs at least $(2+2\beta-\eps)\cdot\OPT$.
Taking the example that corresponds to the larger of the two terms proves the theorem.
\end{proof}

We write $\ALG$ for the cost of the output of the algorithm, and $\eta>0$ is small.
In both examples, the black edges have cost $0$ and form a disconnected forest $H_0$, and every other edge costs its rate for $H=H_0$ (dark red edges) or $1+\eta$ times this rate (green edges, dashed in the figures).
So the algorithm first purchases the black edges, then grows the dual for one unit of time, increasing the dual objective by $D(t)$ (\Cref{lem:dualrate}), and then purchases exactly the dark red edges, which complete a $2$-edge-connected graph.
The reverse delete keeps all edges, as each edge lies in a cut with exactly two purchased edges.
The black and green edges form a feasible solution of cost $(1+\eta)D(t)$, so $D(t)\le\OPT\le(1+\eta)D(t)$, and the dual objective value attained by the algorithm is within a factor $1+\eta$ of $\OPT$.

\FigureDigon

\paragraph{First example (Figure~\ref{fig:digon}).}
Take $k\ge2$ pairs $u_i,v_i$, joined by an edge $u_iv_i$ of cost $0$, and a vertex $g$.
The dark red edges are a parallel copy $s_i$ of every $u_iv_i$, of cost $2$, two copies of $u_1u_i$ for every $i\ge2$, of cost $2+2\beta$, and two copies of $gv_1$, of cost $2+\beta$.
The green edges are $v_iu_{i+1}$ for every $i<k$, of cost $(1+\eta)(2+2\beta)$, and $v_kg$ and $gu_1$, of cost $(1+\eta)(2+\beta)$; with the black edges, they form a Hamiltonian cycle.
Here $H_0$ has $2k$ leaves, one isolated node and $k$ nontrivial components, so $D(t)=k(2+2\beta)+2$ and
\[
\ALG\ =\ k(6+4\beta)-2\beta,\qquad \OPT\ \le\ (1+\eta)\big(k(2+2\beta)+2\big),
\]
and $\ALG/\OPT$ tends to $2+\frac{1}{1+\beta}$ as $k\to\infty$ and $\eta\to0$.

\FigureHub

\paragraph{Second example (Figure~\ref{fig:hub}).}
Take a star with centre $z$ and leaves $l_1,\dots,l_K$, where $K\ge2$, and an edge $p_0p_1$, all of cost $0$.
The dark red edges are $p_0l_j$ for every $j$, of cost $2+2\beta$, and $p_1z$, of cost $1+2\beta$.
The green edges are a second copy of every $zl_j$, of cost $1+\eta$, and $p_0z$ and a second copy of $p_1z$, of cost $(1+\eta)(1+2\beta)$.
Here $H_0$ has $K+2$ leaves and two nontrivial components, so $D(t)=K+2+4\beta$ and
\[
\ALG\ =\ K(2+2\beta)+1+2\beta,\qquad \OPT\ \le\ (1+\eta)(K+2+4\beta),
\]
and $\ALG/\OPT$ tends to $2+2\beta$ as $K\to\infty$ and $\eta\to0$.

\section{Minimal Augmentations of a Forest} \label{sec:minaug}

Our main goal in this section is to prove \Cref{thm:aug}. 
We begin by recalling the theorem setup and relevant notation.   

Throughout this section, we fix a forest $H = (V,F)$, a minimal augmentation $A$ of $H$, and a parameter $\beta \in [0,1/2]$.  
We may assume $|V| \geq 2$ since otherwise the theorem is trivial.  
We refer to edges of $F$ and $A$ as \emph{forest} and \emph{augmentation} edges, respectively. 
For any vertex-set $S \sse V$, we define $d(S)$ and $a(S)$ to be the numbers of forest and augmentation edges, respectively, with exactly one endpoint in $S$. 
For brevity, we use $d(v) = d(\{v\})$ and $a(v) = a(\{v\})$ for any $v \in V$. 
For $S\sse V$, we denote by $H[S]$ the subgraph of $H$ induced by $S$.
By definition of a minimal augmentation, we have $d(S) + a(S) \geq 2$ for any $\emptyset \neq S \ssne V$. 
Moreover, for any augmentation edge $e$, there exists an $\emptyset \neq S_e \ssne V$ such that $e \in \del(S_e)$ and $d(S_e) + a(S_e) = 2$ holds. 
Such an $S_e$ is said to \emph{witness} $e$.

We now define the set of leaves $L \coloneqq \{ u \in V : d(u) = 1 \}$ and the set of isolated vertices $I \coloneqq \{ u \in V : d(u) = 0 \}$ in $H$. 
We refer to the vertices in $L \cup I$ as \emph{active} vertices. 
We also define $\C$ to be the collection of vertex sets of the nontrivial connected components of $H$.
Note that for any $C \in \C$, $d(C) = 0$, and if $C \neq V$, we also have $a(C) \geq 2$. 
For each augmentation edge $e \in A$, we define its \emph{implied cost (or weight)} $w_e$ to be the number of endpoints of $e$ that are active plus $\beta$ times the number of nontrivial components $C \in \C$ crossed by $e$. 
Note that the left-hand side of \Cref{eq:aug} is precisely $w(A) = \sum_{e \in A} w_e$, so proving \Cref{thm:aug} amounts to showing that the implied cost of any minimal augmentation is at most $2(1+\beta) \cdot \bigl( |L| + 2|I| + \frac{|\C|}{1+\beta} - 2 \bigr)$.

\smallskip

\noindent The rest of the section is divided into four subsections:
\begin{itemize}[leftmargin=*, nosep]
    \item In \Cref{subsec:edgetypes}, we classify augmentation edges as \emph{twin} or \emph{mixed}.
    A twin edge $e$ admits a ``twin cut'' witness $S_e$ with $d(S_e)=0$ and $a(S_e)=2$.
    Mixed edges are further classified as \emph{inner} or \emph{cross}, according to whether their endpoints lie in the same or different components of $H$.

    \item In \Cref{subsec:witness}, we carefully choose a minimal witness for each augmentation edge and show that the resulting family of witness sets is \emph{laminar}.
    
    \item In \Cref{subsec:treerep}, we represent this laminar family by a rooted tree and associate each witness with its \emph{exclusive region}, consisting of the vertices in the witness but in none of its children.
    We show that every mixed augmentation edge and its partner forest edge (in the witness) join the exclusive regions of that witness and its parent.

    \item In \Cref{subsec:counting}, we bound the implied cost by separately counting twin edges, incidences of mixed edges at vertices in $L\cup I$, and crossings of mixed edges with respect to $\C$.
\end{itemize}

We use the same conventions in all figures.
Active vertices are represented by black squares, inactive vertices by white circles, and grey dots represent vertices that may be of either type. 
Nontrivial components of the forest are shaded ellipses.
We label every active vertex with $1$ and every nontrivial component with $\beta$, in accordance with the coefficients that appear on the LHS of \eqref{eq:aug}. 
Forest edges are drawn in black and augmentation edges in red. Among augmentation edges, cross edges are drawn as zigzag lines, inner edges as dotted lines, and twin edges as dashed lines; a solid red edge denotes an augmentation edge whose type is irrelevant. 
In the tree of regions, a mixed arc is drawn as a black and a red edge side by side, and a twin arc as red dashed edges.
Coloured outlines and shaded sets mark cuts, witness sets and regions.

\subsection{Classification of augmentation edges} \label{subsec:edgetypes}

Recall that a set $\emptyset \neq S \ssne V$ \emph{witnesses} an augmentation edge $e$ if
$e \in \delta(S)$ and $a(S) + d(S) = 2$.
Since $e \in A \cap \delta(S)$, every witness $S$ of $e$ satisfies either $a(S) = 2$ and $d(S) = 0$, or $a(S) = d(S) = 1$.
We call $S$ a \emph{twin witness} in the former case and a \emph{mixed witness} in the latter.

\begin{definition}[Twin, Mixed, Inner, and Cross Edges] \label{def:edgetypes}
An augmentation edge $e$ is a \emph{twin edge} if it admits a twin witness, and a \emph{mixed edge} otherwise.
A mixed edge is an \emph{inner edge} if its endpoints lie in the same connected component of $H$, and a \emph{cross edge} otherwise.
\end{definition}

We denote the corresponding sets of edges by $\Atw$, $\Amx$, $\Ain$, and $\Acr$, respectively.
Thus, $A= \Atw \sqcup \Amx$ and $\Amx = \Ain \sqcup \Acr$.
Note that every twin witness is a union of connected components of $H$.
Hence, any augmentation edge whose endpoints lie in the same component of $H$ is necessarily an inner edge.
Moreover, every witness of a mixed edge is a mixed witness and therefore has exactly one forest edge in its cut.
\Cref{fig:types} illustrates the different types of augmentation edges. 

\FigureTypes

The above classification allows us to upper bound the implied cost of $A$ by a sum of three simple quantities:
\begin{equation}\label[ineq]{eq:weights}
w(A) \leq (2+2\beta)|\Atw| \  + \ \sum_{e \in \Amx} |V(e) \cap (L \cup I)| \ + \ 2 \beta |\Acr|,
\end{equation}
where $V(e)$ denotes the set of endpoints of $e$. 
Indeed, a twin edge has implied cost at most $2+2\beta$, an inner edge crosses no nontrivial component of $H$ and therefore contributes only through its active endpoints, and a cross edge crosses at most two nontrivial components of $H$.

\subsection{Witness sets and their laminar structure} \label{subsec:witness}

Fix an arbitrary root vertex $r \in V$.
Since $(V,F \cup A)$ is connected, every nonempty cut has a unique representation $\delta(S)$ with $r \notin S$.
For each augmentation edge $e \in A$, we choose a witness $W_e$ avoiding $r$ according to the following rule:
\begin{itemize}[nosep]
\item If $e$ is a twin edge, take $W_e$ inclusion-minimal among all twin witnesses of $e$ that avoid $r$.
\item If $e$ is a mixed edge, take $W_e$ inclusion-minimal among all witnesses of $e$ that avoid $r$.
\end{itemize}
Under either rule, there are one or two inclusion-minimal choices for $W_e$; whenever there are two, we choose one arbitrarily. 

\smallskip

We call $W_e$ the \emph{witness set} of $e$, and refer to $e$ as a \emph{primary edge} of $W_e$.
The resulting \emph{witness family} is $\cW = \{W_e : e \in A\}$.
\Cref{fig:witness} illustrates the selection rule.

\FigureWitness

\begin{restatable}{lemma}{lemwitshape}\label{lem:witshape}
Let $e \in A$ and $W = W_e$.
\begin{enumerate}[label=\upshape(\alph*),nosep]
\item If $e$ is a mixed edge, then $\delta(W) = \{e,\fvar{W}\}$ for some forest edge $\fvar{W}$, and $e$ is the only primary edge of $W$.
We denote this primary edge by $\evar{W}$. 
\item If $e$ is a twin edge, then $W$ is a twin witness and hence a union of connected components of $H$. 
The cut $\delta(W)$ consists of two twin edges, one or both of which are primary edges of $W$.
\item If $S \ssne W$ witnesses $e$, then $e$ is a twin edge and $S$ is a mixed witness.
\end{enumerate}
\end{restatable}

\begin{proof}
Since $W$ witnesses $e$, we have $e \in \delta(W)$ and $d(W)+a(W)=2$.
Moreover, every primary edge of $W$ lies in $\delta(W)$, as $W$ witnesses it.
(a) As $e$ is a mixed edge, it has no twin witness, so $W$ is a mixed witness and $d(W)=a(W)=1$.
Hence $\delta(W)=\{e,\fvar{W}\}$ for some forest edge $\fvar{W}$.
A primary edge of $W$ is an augmentation edge in $\delta(W)$, so it is $e$.
(b) By the choice rule, $W$ is a twin witness, so $d(W)=0$ and $W$ is a union of connected components of $H$.
Hence $\delta(W)$ consists of two augmentation edges, and $W$ is a twin witness of both of them, so both are twin edges.
The primary edges of $W$ lie in $\delta(W)$, and $e$ is one of them.
(c) $S$ avoids $r$, as $S\ssne W$.
If $e$ were a mixed edge, $S$ would be a witness of $e$ that avoids $r$ and lies strictly inside $W$, contradicting the minimality of $W$.
So $e$ is a twin edge.
By the minimality of $W$, $S$ is not a twin witness, so it is a mixed witness.
Finally, by (a) and (b), $d(W_{e'})=1$ for every mixed edge $e'$ and $d(W_{e'})=0$ for every twin edge $e'$, so no witness set has primary edges of both kinds.
\end{proof}

Let $\Wtw$ and $\Wmx$ denote the families of twin and mixed witnesses in $\cW$, respectively, so that $\cW = \Wtw \sqcup \Wmx$.
Since every augmentation edge is a primary edge of exactly one witness set, \Cref{lem:witshape}(a)~and~(b) imply the following. 

\begin{lemma}\label{lem:twin}
We have $|\Amx| = |\Wmx|$ and $|\Atw| \leq 2 |\Wtw|$.
\end{lemma}

The main structural property of the chosen witnesses is their laminarity.
This allows us to represent $\cW \cup \{V\}$ as a rooted tree, which we study in \Cref{subsec:treerep}. 
We prove the following lemma by a standard uncrossing argument. 

\begin{restatable}{lemma}{lemlam}\label{lem:lam}
$\cW$ forms a laminar family. 
\end{restatable}

\begin{proof}
Suppose, for the sake of contradiction, that two witness sets $W, W' \in \cW$ are neither nested nor disjoint. 
Since witnesses avoid $r$, we get that each of the following four derived sets is nonempty:
\[
P \coloneqq W \sm W', \qquad Q \coloneqq W \cap W', \qquad
R \coloneqq W' \sm W, \qquad T \coloneqq V \sm (W \cup W').
\]

To avoid cumbersome notation, we use the shorthand $d'(S)$ to denote $d(S) + a(S)$ for any $S \sse V$. 
Since every $S \in \{P,Q,R,T\}$ is a proper subset of $V$, we get that $d'(S) \geq 2$. 

Since $d'(W) = d'(W') = 2$, submodularity of the cut function and symmetry yield the following tight inequalities:
\begin{align*}
4 = d'(W) + d'(W') &\geq d'(Q) + d'(T) \geq 4,\\
4 = d'(W) + d'(W') &\geq d'(P) + d'(R) \geq 4.
\end{align*}
Consequently, $d'(P) = d'(Q) = d'(R) = d'(T) = 2$ holds.

Let $e$ be a primary edge of $W$, and let $S \in \{P,Q\}$ contain the endpoint of $e$ in $W$. 
Observe that $S \ssne W$ witnesses $e$ since $e \in \delta(S)$ and $d'(S)=2$ holds. 
By \Cref{lem:witshape}(c), $e$ is a twin edge and $S$ is a mixed witness. 
In particular, $W$ is a twin witness. 
Applying the same argument to a primary edge of $W'$, using $Q$ and $R$, shows that $W'$ is also a twin witness.

Since $W$ and $W'$ are twin witnesses, both are unions of connected components of $H$.
Their difference $P$ and intersection $Q$ are therefore also unions of connected components of $H$.
Since $S \in \{P,Q\}$, we have $d(S)=0$, contradicting that $S$ is a mixed witness. 
\end{proof}

\subsection{Tree representation of witness sets and their exclusive regions} \label{subsec:treerep}

Since $\cW$ is laminar, we can represent $\cW \cup \{V\}$ as a rooted tree, which we call the \emph{witness tree}, with root $V$. 
For each $W \in \cW$, its \emph{parent} $\parent(W)$ is the smallest set in $\cW \cup \{V\}$ that strictly contains $W$. 
For $W \in \cW \cup \{V\}$, we denote by $\children(W)$ the collection of its \emph{children}, namely the sets whose parent is $W$.
Note that $\children(W)$ may be empty.  
Next, for any $W \in \cW \cup \{V\}$, we define its \emph{exclusive region}, or simply its \emph{region}, as $\region(W) \coloneqq W \sm \bigcup_{W' \in \children(W)} W'$. 
The regions are pairwise disjoint by laminarity (\Cref{lem:lam}) and cover $V$. 
Note that the \emph{root region} $\region(V)$ contains $r$, since every witness set avoids $r$. 

\begin{lemma}\label{lem:nonempty}
For every $W \in \cW \cup \{V\}$, $\region(W) \neq \emptyset$. 
More precisely, if $e$ is a primary edge of $W$, then the endpoint of $e$ in $W$ lies in $\region(W)$. 
\end{lemma}

\begin{proof}
Since $\region(V)$ contains $r$, it suffices to consider $W \in \cW$.
Let $e$ be a primary edge of $W$. 
If the endpoint of $e$ in $W$ is not in $\region(W)$, it lies in a child $W' \in \cW$ of $W$. 
The other endpoint of $e$ is outside $W$, so $e \in \delta(W')$. 
Since $W' \ssne W$ witnesses $e$, \Cref{lem:witshape}(c) implies that $e$ is a twin edge and $W'$ is a mixed witness. 
Since $e$ is a primary edge of $W$, it cannot be a primary edge of $W'$. 
So, $\delta(W')$ contains two distinct augmentation edges: $e$ and a primary edge $e'$ of $W'$. 
This contradicts our earlier conclusion that $W'$ is a mixed witness. See \Cref{fig:nonempty} for an illustration of the proof. 
\end{proof}

\FigureNonempty

Since the regions are nonempty and pairwise disjoint, each region $X$ corresponds to a unique set $\witness(X) \in \cW \cup \{V\}$ such that $X = \region(\witness(X))$.
Relabeling each node $W$ of the witness tree by $\region(W)$ gives the \emph{tree of regions} $\cT$, rooted at $\region(V)$.
We denote the parent of a nonroot region $X$ by $\parent(X)$ and the children of any region $X$ by $\children(X)$.
In particular, $\parent(X) = \region(\parent(\witness(X)))$ for every nonroot region $X$.

We orient each edge of $\cT$ towards the root. 
For a nonroot region $X$, let $W = \witness(X)$.
We classify the $\cT$-arc going from $X$ to $\parent(X)$ as follows:
\begin{itemize}[nosep]
\item If $W \in \Wmx$, the arc is a \emph{mixed arc}, and we assign to it the two edges in $\delta(W)$: the augmentation edge $\evar{W} \in A$ and the forest edge $\fvar{W} \in F$.
It is an \emph{inner arc} or a \emph{cross arc} according to whether $\evar{W}$ is an inner or a cross edge.
\item If $W \in \Wtw$, the arc is a \emph{twin arc}, and we assign the primary edges of $W$ to it.
\end{itemize}

Observe that every augmentation edge is assigned to exactly one $\cT$-arc, namely the arc corresponding to its chosen witness set. 
Clearly, the tree $\cT$ has $|\cW|+1$ nodes and $|\cW|$ arcs, of which $|\Wmx|$ are mixed and $|\Wtw|$ are twin. 
\Cref{fig:regions} shows the tree of regions for the graph in \Cref{fig:families}.

\FigureFamilies

\FigureRegions

The following lemma shows that the two edges assigned to each mixed arc join its incident regions, and that every forest edge joining distinct regions is assigned to exactly one mixed arc. 

\begin{lemma}\label{lem:arcs}
\begin{enumerate}[label=\upshape(\roman*),nosep]
\item For every mixed arc $(X,\parent(X))$ of $\cT$, both edges assigned to it have one endpoint in $X$ and the other in $\parent(X)$.
\item Every forest edge joining distinct regions is assigned to exactly one mixed arc of $\cT$.
\end{enumerate}
\end{lemma}

\begin{proof}
For each $U \in \cW$, the union of the regions in the subtree rooted at $\region(U)$ is $U$.
Thus, for $U \in \cW$ and $g \in F \cup A$, we have $g \in \delta(U)$ if and only if the arc $(\region(U),\region(\parent(U)))$ lies on the unique undirected path in $\cT$ between the regions containing the endpoints of $g$.

Fix $W \in \Wmx$. 
Let $e = \evar{W}$ and $f = \fvar{W}$, so that $\delta(W) = \{e,f\}$.
We show that neither edge belongs to the cut of any other chosen witness.
Let $W' \in \cW \sm \{W\}$, and let $e'$ be a primary edge of $W'$.
Since every augmentation edge has exactly one chosen witness set, $e' \neq e$.
If $e \in \delta(W')$, then $\delta(W') = \{e,e'\}$, making $W'$ a twin witness of $e$, contrary to $e$ being mixed.

If $f \in \delta(W')$, then $\delta(W') = \{f,e'\}$.
Observe that $\delta(W \triangle W') = \delta(W) \triangle \delta(W')$, which is $\{e,e'\}$.
Since $W \neq W'$ and both avoid $r$, the set $W \triangle W'$ is nonempty and proper.
It is therefore a twin witness of $e$, again a contradiction.
Consequently, each of $e$ and $f$ belongs to exactly one chosen witness cut, namely $\delta(W)$.
By the path observation above, the regions containing its endpoints are joined by the single arc corresponding to $W$.
Thus, both edges join $\region(W)$ and $\region(\parent(W))$, proving (i).

For (ii), now let $f \in F$ join two distinct regions.
The undirected path between them in $\cT$ contains an arc $(\region(W),\region(\parent(W)))$ for some $W \in \cW$, so $f \in \delta(W)$.
Since $W$ is a chosen witness and $f$ is a forest edge, $W$ is mixed and $f = \fvar{W}$.
Thus, $f$ is assigned to this mixed arc.
The preceding argument shows that $f$ belongs to no other chosen witness cut, so it is assigned to no other arc. 
\end{proof}

\begin{lemma}\label{lem:exits}
Every set $S\sse V$ satisfies $|L\cap S|+2|I\cap S|+d(S)\ge2\kappa(H[S])$.
\end{lemma}

\begin{proof}
Let $q=\kappa(H[S])$ be the number of connected components of $H[S]$. 
We start with the case that $H[S]$ is connected (Figure~\ref{fig:exits}). Its vertices and the forest edges inside it are connected and have no cycle, so there
are $|S|-1$ such edges. Adding up the degrees in $H$ of the vertices of $S$ counts every edge inside $S$ twice and every forest edge in $\del(S)$ once, so
$\sum_{v\in S}d(v)=2(|S|-1)+d(S)$. We rewrite this as
\[
\textstyle\sum_{v\in S}\big(2-d(v)\big)\ =\ 2-d(S) .
\]
In this sum, a vertex of degree $1$ is a leaf and contributes $1$. A vertex of degree $0$ is a component on its own, so it is isolated,
and it contributes $2$.  Every other vertex contributes at most $0$. Hence $|L\cap S|+2|I\cap S|\ge2-d(S)$.

Now let $S$ be any set, and let $S_1,\dots,S_q$ be the vertex sets of the components of $H[S]$. Since no forest edge joins two of them,
every forest edge in $\delta(S_i)$ lies in $\delta(S)$, and every forest edge in $\delta(S)$ lies in $\delta(S_i)$ for exactly one $i$. So
$d(S_1)+\dots+d(S_q)=d(S)$, and adding the inequality above over $S_1,\dots,S_q$ gives 
\[
|L\cap S|+2|I\cap S|\ge2q-d(S). \qedhere 
\]
\end{proof}

\FigureExits

The correspondence in \Cref{lem:arcs} allows us to count mixed-edge endpoints in a region by counting forest edges with exactly one endpoint in that region.

\begin{restatable}{lemma}{lemdeg}\label{lem:deg}
For every region $X$, the number of mixed arcs of $\cT$ incident to $X$, counting both incoming and outgoing arcs, equals both $d(X)$ and the number of endpoints of mixed edges in $X$.
If $X$ contains no active vertex, i.e., $X \cap (L \cup I) = \emptyset$, then $d(X) \geq 2$. 
\end{restatable}

\begin{proof}
Take a mixed arc at $X$, say the arc of the region $X'$, so that $X$ is $X'$ or $\parent(X')$, and let $W'=\witness(X')$. By
Lemma~\ref{lem:arcs}(i), both of its edges $\fvar{W'}$ and $\evar{W'}$ join $X'$ and $\parent(X')$, so each of them has exactly one endpoint in $X$
(Figure~\ref{fig:deg}). In particular $\fvar{W'}$ lies in $\delta(X)$.

Conversely, every forest edge in $\delta(X)$ and every mixed edge with an endpoint in $X$ come from such an arc. A forest edge in $\delta(X)$ joins $X$ to another region, so by
Lemma~\ref{lem:arcs}(ii) it is $\fvar{W'}$ for exactly one $W'\in\Wmx$, and the arc of $\region(W')$ joins the two regions of its endpoints, one
of which is $X$. A mixed edge $e$ is $\evar{W'}$ for its witness set $W'=W_e$ (Lemma~\ref{lem:witshape}(a)), and it joins $\region(W')$ and
$\region(\parent(W'))$, so if $e$ has an endpoint in $X$, then the arc of $\region(W')$ is at $X$. Different arcs carry different edges, so the
number of mixed arcs at $X$ is both $d(X)$ and the number of endpoints of mixed edges in $X$.

Finally, suppose that $X$ contains no active vertex. Then $X$ is nonempty (Lemma~\ref{lem:nonempty}), so $\kappa(H[X])\ge1$, and
Lemma~\ref{lem:exits} gives $d(X)\ge2\kappa(H[X])\ge2$.
\end{proof}

\FigureDeg

Finally, consider the forest obtained from $\cT$ by deleting all twin arcs and ignoring orientations.
We call its connected components \emph{clusters} and identify each cluster with the union of its regions.
\Cref{fig:clustertree} illustrates this decomposition.

\FigureClusterTree

\begin{restatable}{lemma}{lemgroups}\label{lem:groups}
There are exactly $|\Wtw|+1$ clusters, and the twin arcs connect them into a tree.
Every cluster is a union of connected components of $H$.
In particular, $H$ has at least $|\Wtw|+1$ connected components.
\end{restatable}

\begin{proof}
Every set $W\in\cW$ gives exactly one arc of the tree of regions, between its region $\region(W)$ and the region $\region(\parent(W))$
of its parent, and this arc is a twin arc if $W\in\Wtw$ and a mixed arc if $W\in\Wmx$. So the tree of regions has $|\cW|+1$ nodes, the
regions, and $|\cW|$ arcs, of which $|\Wtw|$ are twin arcs and $|\Wmx|=|\cW|-|\Wtw|$ are mixed arcs. The clusters are the connected
components of the graph on the regions whose edges are the mixed arcs.

For every $W\in\Wtw$, its \emph{representative} is the set of its primary edges: one twin edge of $\delta(W)$, or both edges of
$\delta(W)$ if both have $W$ as their witness set. Every twin witness set has its own representative, and no other witness set has it,
since every augmentation edge is a primary edge of exactly one witness set. The twin arc of $W$ carries exactly the representative of
$W$, so deleting the twin arcs means deleting the representatives, and we delete exactly as many representatives as there are twin
witness sets, namely $|\Wtw|$. Deleting an arc from a forest splits the component that contains it into two: since there is no cycle, the
two endpoints of the deleted arc can no longer reach each other, and the other components do not change. So every deletion adds exactly one
component, and as the tree has one component, the number of clusters is the number of deleted representatives plus one, that is,
$|\Wtw|+1$. Equivalently, the mixed arcs form a forest with $|\cW|+1$ nodes and $|\cW|-|\Wtw|$ arcs, and a forest has
as many components as nodes minus arcs: $(|\cW|+1)-(|\cW|-|\Wtw|)=|\Wtw|+1$. A forest edge either lies inside
a region, or it joins two regions; in the second case it is $\fvar{W}$ for some $W\in\Wmx$, and it joins the two regions of the arc of
$\region(W)$ (Lemma~\ref{lem:arcs}), which lie in the same cluster. So no forest edge joins two clusters, and every component of $H$ lies
inside one cluster. The regions are nonempty (Lemma~\ref{lem:nonempty}) and disjoint, so the clusters are nonempty and disjoint as
well. Each of them is a union of whole components and contains at least one, so $H$ has at least $|\Wtw|+1$ components.
\end{proof}

\subsection{Counting incidences} \label{subsec:counting}

To bound the right side of \eqref{eq:weights}, it remains to count three quantities: the number of twin edges, the number of endpoints of mixed edges at active vertices, and the number of cross edges.

The twin edges are counted by \Cref{lem:twin}, which gives $|\Atw|\le2|\Wtw|$.

Next, we bound the number of endpoints of mixed edges at active vertices by counting the mixed arcs at every region.
We call a region \emph{active} if it contains an active vertex, and \emph{inactive} otherwise.

\begin{lemma}\label{lem:cact}
$\sum_{e\in\Amx}|V(e)\cap(L\cup I)|\le2\bigl(|L|+|I|-|\Wtw|\bigr)-2$.
\end{lemma}
 
\begin{proof}
Let $\nactive$ be the number of active regions.
The regions are disjoint and each active region contains an active vertex, so $\nactive\le|L|+|I|$.
By Lemma~\ref{lem:deg}, $d(X)$ is the number of mixed arcs at a region $X$.
Every mixed arc joins two regions, and there are $|\Wmx|=|\cW|-|\Wtw|$ mixed arcs, so $\sum_Xd(X)=2|\cW|-2|\Wtw|$.
There are $|\cW|+1-\nactive$ inactive regions, and each of them has $d(X)\ge2$ by Lemma~\ref{lem:deg}.
Hence $\sum_{X\text{ active}}d(X)\le2|\cW|-2|\Wtw|-2(|\cW|+1-\nactive)=2\nactive-2-2|\Wtw|$.
Every endpoint of a mixed edge at an active vertex lies in an active region $X$, and $X$ contains exactly $d(X)$ endpoints of mixed edges, again by Lemma~\ref{lem:deg}.
So the mixed edges have at most $2\nactive-2-2|\Wtw|\le2\bigl(|L|+|I|-|\Wtw|\bigr)-2$ endpoints at active vertices.
\end{proof}

Recall that $H[S]$ denotes the subgraph of $H$ induced by $S$ for any $S\sse V$. 
It remains to count the cross edges, which we do one cluster at a time.
Fix a cluster $C$.
By Lemma~\ref{lem:groups}, $C$ is a union of whole components of $H$, so $\kappa(H[C])$ is the number of components of $H$ in $C$, and the regions of $C$ with the mixed arcs between them form a tree.
The \emph{cross edges of $C$} are the cross edges carried by the cross arcs of this tree.
Each cross arc carries exactly one cross edge, and a tree has one arc fewer than nodes, so
\begin{equation}\label{eq:crossG}
\#\,\text{cross edges of }C\ =\ \#\,\text{regions of }C\ -\ 1\ -\ \#\,\text{inner arcs of }C .
\end{equation}

A region $X$ is \emph{connected} if $H[X]$ is connected, and \emph{disconnected} otherwise (Figure~\ref{fig:clusters}).
\FigureClusters

\begin{lemma}\label{lem:junctions}
For every cluster $C$,
\[
\sum_{X}\bigl(\kappa(H[X])-1\bigr)=\kappa(H[C])-1,
\]
where the sum ranges over all regions $X$ of $C$.
Consequently, $C$ has at most $\kappa(H[C])-1$ disconnected regions.
\end{lemma}

\begin{proof}
$C$ is the union of the $\kappa(H[C])$ components of $H[C]$. In each of them, delete the forest edges that join different regions.
What remains of the component falls into connected parts, and each part lies inside one region. Two vertices of a region that are joined by
forest edges inside the region are in the same part, so these parts are exactly the vertex sets of the components of the graphs $H[X]$. Deleting edges from a
component of $H$ leaves one more part than the number of deleted edges, since a component is connected and has no cycle. Adding over
the components, $\sum_X\kappa(H[X])$ is the number of deleted edges plus $\kappa(H[C])$. By Lemma~\ref{lem:arcs}, the deleted edges are
exactly the forest edges of the arcs of $C$, one for each arc, and $C$ has one arc fewer than regions. So
$\sum_X\big(\kappa(H[X])-1\big)=\kappa(H[C])-1$. Every disconnected region adds at least $1$ to this sum.
\end{proof}

\begin{lemma}\label{lem:transit}
For every cross arc, at least one of the two regions that it joins is disconnected.
\end{lemma}

\begin{proof}
By Lemma~\ref{lem:arcs}(i), the forest edge and the cross edge of the arc both join its two regions (Figure~\ref{fig:transit}).
If one of the two regions is connected, all its vertices belong to one component of $H$.
Then the other endpoint of the forest edge lies in this component, and the other endpoint of the cross edge does not.
Both lie in the second region, which is therefore disconnected.
\end{proof}
\FigureTransit

\begin{lemma}\label{lem:fewtransit}
If $\kappa(H[C])\ge2$, then $C$ has at most $\kappa(H[C])-2$ inactive connected regions plus one per inner arc of $C$.
\end{lemma}

\begin{proof}
Throughout this proof, parent--child relations are restricted to the tree
of mixed arcs within $C$. 
The claim is trivial when $C$ has no inactive connected regions, so suppose otherwise. 
The \emph{highest region} of $C$ is the region of $C$ whose parent region is not in $C$, or the root region. 
The arcs incident to a region join it to its parent and child regions, and the \emph{children} of a region of $C$ are the regions of $C$ whose parent region it is (Figure~\ref{fig:fewtransit}).
An inactive connected region has at least two mixed arcs (Lemma~\ref{lem:deg}) incident to it. 
Since at most one such arc goes to the parent region, there must exist a child. 

Put one token on every disconnected region of $C$ and one on every inner arc of $C$.
Every inactive connected region $X$ chooses a child $X'$ and takes the token of the arc between them if it is an inner arc, and the token of $X'$ if it is a cross arc. We know that this token exists, as $X'$ is then disconnected by Lemma~\ref{lem:transit}.

We now show that one token is never taken.
If the highest region is disconnected, nobody takes its token, as its parent region is not in $C$.
Otherwise it is connected, and it has a child in $C$, since $C$ contains an inactive connected region and its child.
If it is inactive, all its mixed arcs join it to children within $C$, so it has at least one child $X'$ that it did not choose. 
If it is active, it takes no token, and $X'$ can be any child. 
So in both cases, either the token of $X'$ or the token of the arc to $X'$ exists and nobody takes it. 
Hence the number of inactive connected regions is at most the number of disconnected regions plus the number of inner arcs, minus one, and Lemma~\ref{lem:junctions} gives the claim.
\end{proof}
\FigureFewTransit

\begin{lemma}\label{lem:c3group}
Every cluster $C$ has at most $|L\cap C|+2|I\cap C|+2\kappa(H[C])-4$ cross edges.
In total, $|\Acr|\le|L|+2|I|+2\kappa(H)-4-4|\Wtw|$.
\end{lemma}

\begin{proof}
If $\kappa(H[C])=1$, then $C$ is a single component of $H$, so it has no cross edges by Lemma~\ref{lem:arcs}(i), and the bound is at least $0$, as $C$ is an isolated vertex or has at least two leaves.
Now let $\kappa(H[C])\ge2$.
At most $\kappa(H[C])-1$ regions of $C$ are disconnected by Lemma~\ref{lem:junctions}.
At most $|L\cap C|+|I\cap C|$ regions are active connected, since each contains an active vertex.
By Lemma~\ref{lem:fewtransit}, at most $\kappa(H[C])-2$ plus the number of inner arcs are inactive connected.
By \eqref{eq:crossG}, we get that $C$ has at most $|L\cap C|+|I\cap C|+2\kappa(H[C])-4$ cross edges, which is what we wanted.

For the total, every cross edge and every component of $H$ belong to exactly one of the $|\Wtw|+1$ clusters (Lemma~\ref{lem:groups}), so adding the first bound over the clusters gives the second.
\end{proof}

\begin{proof}[Proof of Theorem~\ref{thm:aug}]\phantomsection\label{sec:end}
The left side of \Cref{thm:aug} is $w(A)$, and we bound the three terms of \eqref{eq:weights} one by one.
There are $|\Atw|\le2|\Wtw|$ twin edges, by \Cref{lem:twin}.
By Lemma~\ref{lem:cact}, the mixed edges have at most $2(|L|+|I|-|\Wtw|)-2$ endpoints at active vertices.
By Lemma~\ref{lem:c3group}, there are $|\Acr|\le|L|+2|I|+2\kappa(H)-4-4|\Wtw|$ cross edges.
Plugging these three bounds into \eqref{eq:weights}, we obtain
\begin{align*}
w(A)\ &\le\ (2+2\beta)\cdot2|\Wtw|+2\big(|L|+|I|-|\Wtw|\big)-2\\
&\qquad+2\beta\big(|L|+2|I|+2\kappa(H)-4-4|\Wtw|\big)\\
&=\ (2+2\beta)\big(|L|+2|I|\big)+2\big(\kappa(H)-|I|\big)-4-4\beta\\
&\qquad-(2-4\beta)\big(\kappa(H)-1-|\Wtw|\big).
\end{align*}
Finally, $H$ has at least $|\Wtw|+1$ components by Lemma~\ref{lem:groups}, and $\beta\le\frac12$, so the last term is at most $0$.
What remains is the right side of \eqref{eq:aug}, as $|\C|=\kappa(H)-|I|$.
\end{proof}

\section*{AI Disclosure} \label{sec:aidisclosure}

No AI tools were used to obtain the mathematical results in this work.
The authors used ChatGPT (OpenAI) to improve grammar and exposition, to obtain feedback on draft proofs, and to assist with drawing the figures.

\printbibliography

\end{document}